\documentclass[sigconf]{acmart}
\def\BibTeX{{\rm B\kern-.05em{\sc i\kern-.025em b}\kern-.08emT\kern-.1667em\lower.7ex\hbox{E}\kern-.125emX}}
    
\usepackage{nicefrac}
\usepackage{siunitx}
\usepackage{array,framed}
\usepackage{booktabs}
\usepackage{
  color,
  float,
  epsfig,
  wrapfig,
  graphics,
  graphicx,
  subcaption
}
\usepackage{textcomp}
\usepackage{setspace}
\usepackage{latexsym,fancyhdr,url}
\usepackage{enumerate}
\usepackage{algorithm2e}
\usepackage{algpseudocode}
\usepackage{graphics}
\usepackage{xparse} % argument parsing -- \edist
\usepackage{xspace}
\usepackage{multirow}
\usepackage{csvsimple}
\usepackage{balance}
\usepackage{
  tikz,
  pgfplots,
  pgfplotstable
}
\usepackage{hyperref}
\hypersetup{
    colorlinks=true,
    linkcolor=blue,
    filecolor=magenta,      
    urlcolor=cyan,
    pdftitle={Overleaf Example},
    pdfpagemode=FullScreen,
    }

\usetikzlibrary{
  shapes.geometric,
  arrows,
  external,
  pgfplots.groupplots,
  matrix
}

\pgfplotsset{compat=1.9}
\usepackage{mathtools,}

\DeclareMathAlphabet{\mathcal}{OMS}{cmsy}{m}{n}
\DeclareGraphicsExtensions{%
    .png,.PNG,%
    .pdf,.PDF,%
    .jpg,.mps,.jpeg,.jbig2,.jb2,.JPG,.JPEG,.JBIG2,.JB2}

\usepackage{xparse}
\newcommand{\bnm}{\begin{newmath}}
\newcommand{\enm}{\end{newmath}}

\newcommand{\bea}{\begin{eqnarray*}}%
\newcommand{\eea}{\end{eqnarray*}}%

\newcommand{\bne}{\begin{newequation}}
\newcommand{\ene}{\end{newequation}}

\newcommand{\bal}{\begin{newalign}}
\newcommand{\eal}{\end{newalign}}

\newenvironment{newalign}{\begin{align}%
\setlength{\abovedisplayskip}{4pt}%
\setlength{\belowdisplayskip}{4pt}%
\setlength{\abovedisplayshortskip}{6pt}%
\setlength{\belowdisplayshortskip}{6pt} }{\end{align}}

\newenvironment{newmath}{\begin{displaymath}%
\setlength{\abovedisplayskip}{4pt}%
\setlength{\belowdisplayskip}{4pt}%
\setlength{\abovedisplayshortskip}{6pt}%
\setlength{\belowdisplayshortskip}{6pt} }{\end{displaymath}}

\newenvironment{newequation}{\begin{equation}%
\setlength{\abovedisplayskip}{4pt}%
\setlength{\belowdisplayskip}{4pt}%
\setlength{\abovedisplayshortskip}{6pt}%
\setlength{\belowdisplayshortskip}{6pt} }{\end{equation}}

\newcounter{ctr}

\newcounter{mytable}
\def\mytable{\begin{centering}\refstepcounter{mytable}}
\def\endmytable{\end{centering}}

\newcounter{myfig}
\def\myfig{\begin{centering}\refstepcounter{myfig}}
\def\endmyfig{\end{centering}}

\newlength{\saveparindent}
\newlength{\saveparskip}
\newcommand{\E}{{\rm I\kern-.3em E}}

\renewcommand{\eqref}[1]{\mbox{Equation~(\ref{#1})}}

\def \part {part}

\renewcommand{\paragraph}[1]{\vspace*{6pt}\noindent\textbf{#1}\;}

\def \blackslug{\hbox{\hskip 1pt \vrule width 4pt height 8pt
    depth 1.5pt \hskip 1pt}}
\def \qed{\quad\blackslug\lower 8.5pt\null\par}
\newcounter{mynote}[section]

\newcommand\ignore[1]{}

\newcounter{rcnote}[section]

\newcounter{mrnote}[section]

\newcounter{fknote}[section]

\newcounter{anote}[section]

\DeclareMathSymbol{\mlq}{\mathord}{operators}{``}
\DeclareMathSymbol{\mrq}{\mathord}{operators}{`'}

\newcommand{\rhf}[2]{R_{f, \gamma}}

\DeclareDocumentCommand{\edist}{o o}{
  \ensuremath{
    \IfNoValueTF{#1}{{d}}{{\sf d}(#1,#2)}
  }
}

\newcommand{\olrk}[1]{\ifx\nursymbol#1\else\!\!\mskip4.5mu plus 0.5mu\left(\mskip0.5mu plus0.5mu #1\mskip1.5mu plus0.5mu \right)\fi}

\NewDocumentCommand{\indseq}{ O{1} O{r} }{{#1}\ldots {#2}}

\usepackage{fontawesome}
\usepackage{pifont}
\newcommand{\mcrot}[4]{\multicolumn{#1}{#2}{\rlap{\rotatebox{#3}{#4}~}}} 

\usepackage[most]{tcolorbox}
\newcommand*\emptycirc[1][1ex]{\tikz\draw[thick] (0,0) circle (#1);} 
\newcommand*\halfcirc[1][1ex]{%
  \begin{tikzpicture}
  \draw[fill] (0,0)-- (90:#1) arc (90:270:#1) -- cycle ;
  \draw[thick] (0,0) circle (#1);
  \end{tikzpicture}}
\newcommand*\fullcirc[1][1ex]{\tikz\fill (0,0) circle (#1);}

\setcopyright{none}
\renewcommand\footnotetextcopyrightpermission[1]{} % removes footnote 
\begin{document}
%\fontfamily{lmr}\selectfont
% \def\thetitle{A Practical Way to Generate Strong Keys from Noisy Data}
\fancyhead{}
\def\thetitle{False Information, Bots and Malicious Campaigns: \\
   Demystifying Elements of Social Media Manipulations}
\title{\thetitle}

\author{Mohammad Majid Akhtar}
\affiliation{\small{University of New South Wales}}

\author{Rahat Masood}
\affiliation{\small{University of New South Wales}}

\author{Muhammad Ikram}
\affiliation{\small{Macquarie University}}

\author{Salil S. Kanhere}
\affiliation{\small{University of New South Wales}}

\date{}

\begin{abstract}
  % Abstractly this is an ACM CCS Template. Keep it short and simple, highlight
  % the main problem and give your punch line contributions. For example,

  % Setting up the ACM CCS template is non-trivial. This is a document to help you
  % get started with ACM CCS template over Overleaf quickly. I also provide some
  % macros in the \texttt{defs.tex} file, that can be helpful for new writers.

The rapid spread of false information and persistent manipulation attacks on online social networks (OSNs), often for political, ideological, or financial gain, has affected the openness of OSNs. While researchers from various disciplines have investigated different manipulation-triggering elements of OSNs (such as understanding information diffusion on OSNs or detecting automated behavior of accounts), these works have not been consolidated to present a comprehensive overview of the interconnections among these elements. Notably, user psychology, the prevalence of bots, and their tactics in relation to false information detection have been overlooked in previous research. 

To address this research gap, this paper synthesizes insights from various disciplines to provide a comprehensive analysis of the manipulation landscape. By integrating the primary elements of social media manipulation (SMM), including false information, bots, and malicious campaigns, we extensively examine each SMM element. Through a systematic investigation of prior research, we identify commonalities, highlight existing gaps, and extract valuable insights in the field. 

Our findings underscore the urgent need for interdisciplinary research to effectively combat social media manipulations, and our systematization can guide future research efforts and assist OSN providers in ensuring the safety and integrity of their platforms.

\end{abstract}

\maketitle
\keywords{LaTeX template, ACM CCS, ACM}

% Section I
\section{Introduction}
\label{sec:intro}
Social media and OSNs
are used for various purposes, including communication, content creation, sharing, and community building. An estimated 4.74 billion users spent an average of two hours daily on social networks in 2022, and this trend is expected to rise in the future~\cite{socialmediauser,socialmediausage}. 
The open nature of social networks plays a vital role in building strong ties and relationships (also called ``connections'') between online users~\cite{fuchs2021social}, who can quickly share valuable and essential information within their circles on any topic or event. For example, approximately 16 million text messages and 347 thousand tweets are sent every minute \cite{socialmediacontent}.

However, social media's inherent openness---the ease of sharing unverified information with many users makes social media vulnerable to false information (or misinformation) spread by fake accounts through malicious campaigns. In particular, false information is untrue or misleading information that humans spread intentionally or unintentionally. Besides, automated software programs, known as {\it bots}, can mimic human behavior online, such as posting on social media or writing comments, thus aiding the spreading of fake information on OSNs. Likewise, malicious campaigns are coordinated efforts to spread false information or manipulate public opinion, often for political, ideological, or financial gain \cite{culloty2021disinformation}.
% , depicted in Figure \ref{fig:categories}. 
These three co-exist such that ``false information'' is ``what'' gets shared, ``bots'' are agents/entities who typically spread ``false information'', and ``malicious campaigns'' are targeted initiatives launched by an adversary. Collectively, these three {\it elements} create a new threat to OSNs: \textit{social media manipulations} (SMMs).

% \begin{figure}[h]
% \centering
% \includegraphics[width=6.5cm]{images/campaign_categories (2).pdf}
% \caption{Impact of Malicious Campaigns on various themes}
% \label{fig:categories}
% \end{figure}

With the proliferation of SMM, various events have observed traces of manipulation attempts with differing impact levels. Many such incidents include worldwide events such as the COVID-19 pandemic ~\cite{antivaccine}, nationwide events such as various elections \cite{uselection,taiwan,french}, and heaps of misinformation, phishing, spam content targeted at individual-level on OSNs. As a result, this leads to confusion and doubts among the OSN users, ultimately causing the erosion of trust in the platform. 

Several studies have explored and reviewed the research on false information \cite{10.1145/3137597.3137600,doi:10.1126/science.aap9559}, bots \cite{cresci2020detecting,botornot}, and campaigns \cite{9518390,starbird2019disinformation}, and have provided definitions for related terms and examined their impact on various aspects of society~\cite{shao2018spread,Lee_Eoff_Caverlee_2021}. For instance, existing surveys on false information detection aim to highlight aspects such as early detection, multimodal detection, and explanatory detection \cite{10.1145/3393880}. At the same time, bot detection surveys emphasize detecting automated behavior irrespective of malicious activity bots perform. On the other hand, studies \cite{10.1145/3393880,https://doi.org/10.48550/arxiv.1804.08559} on campaign detection emphasize quantifying the impact of information on humans rather than identifying the malicious strategies used by fake accounts \cite{perspective}. However, we find that most of these studies lack an understanding of their relationship with one another. For example, studies focusing on false information do not explore the actors involved in propagating false information, and studies on bot detection do not focus on how bots leverage different strategies and behaviors, such as campaigns. In contrast, while we discuss the three SMM elements and their challenges, our primary focus is understanding each element and its relationship with the other elements in the OSN landscape, which has not been discussed comprehensively in existing works. We hope that highlighting the intent and strategies used by bots in malicious campaigns inspires the research community to build robust solutions against SMMs.

\begin{figure*}[h]
    \centering
    \includegraphics[width=\linewidth]{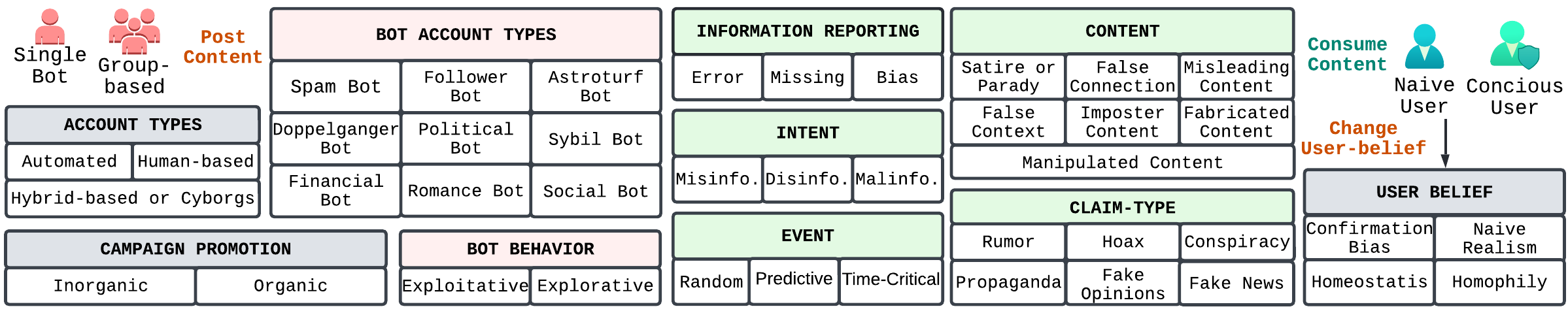}
    % \vspace{-1em}
    % \setlength{\belowcaptionskip}{-23pt}
    \caption{Manipulation landscape and its elements.}
    \label{fig:informationdisorder}
    \vspace{+0.2cm}
\end{figure*}

\textit{Filling the Gap.} By analyzing the current literature on SMMs, we found that (1) researchers have separately and independently investigated the three elements of SMMs: false information, bots, and malicious campaigns, and no emphasis has been placed on assessing the relationship between them; 
and (2) compared to false information and bot detection works, there is a lack of work on the systematization and unification of the knowledge on malicious campaigns. 
Our work distinguishes itself from prior research by examining all three elements of SMMs. 
In particular, we systematically examine the objectives, intentions, and strategies employed by bots and fake accounts in malicious campaigns
% \footnote{In this paper, we have used misinformation, false information, and disinformation interchangeably. Disinformation campaigns, malicious campaigns, influence operations, 
% strategic information operations, and coordinated activity are used interchangeably.} 
that have only been partially explored in previous literature. We
highlight the need for increased efforts to detect coordinated malicious activities. 
Thus, our contribution to the field is threefold: (i) we explore the interplay of different SMM elements by dissecting and organizing existing findings, highlighting commonalities, gaps, and takeaways in the field, (ii) we draw on insights from other disciplines to provide a fresh perspective on the OSN manipulation landscape, and (iii) {discuss} a framework for assessing which intent and strategy are used by an adversary in SMM. Finally, we combine the various SMM factors to raise interesting questions for future research.

%%% Local Variables:
%%% mode: latex
%%% TeX-master: "main"
%%% End:

%  LocalWords:  biometrics cryptographic parallelized lossy
    % basic introduction
% \vspace{-1em}
\section{Systematic Review Protocol}
\label{sec:protocol}
{
As defined Section~\ref{sec:intro}, in this research work, we cover three types of SMMs: false information, bots, and malicious campaigns. We searched and selected papers following a systematic literature review (SLR) process. We form search words (as elaborated in the Appendix \ref{appendix:systematic}) for the three respective SMMs to query separately from databases such as Scopus, IEEE Digital Library, ACM Library, and Google Scholar. To query papers from these databases, we combine keywords in four dimensions: purpose, type of SMM, method, and used platform. 

Since SMM is a recent threat in the context of OSN, we kept the timeline for search queries from 2015 to 2022 to cover the past half-decade and the current work in SMM. In some places, we use papers before 2015 (publication year) to support a few definitions and to provide the necessary background. We first exclude papers that do not pertain to the chosen SMM factors or are not written in English. Next, we analyze papers by reading their title and abstract to filter out the most notable works. Next, we proceed to our second-level reading process using a set of criteria. The criteria include looking into aspects such as the category of false information type (fake news, rumor, hoax, propaganda, phishing), as all are commonly used definitions for false information. Other criteria include the OSN platform (such as Twitter (rebranded as $\mathbb{X}$) and Facebook), the method or technique used, and the most prevalent features in the detection process. The bot type categories (spambot, social bot, follower bot, scam bot), with other aforementioned criteria, define the analysis criteria for bot detection-related articles. Likewise, we analyze malicious campaign detection papers on the defined criteria. This results in 32 papers for analysis in false information detection, 35 in bot detection, and 15 in malicious campaign detection. We note that limited work exists in malicious campaign detection.}

\vspace{-1em}
\section{OSNs Manipulations Landscape}
\label{sub:osnmanipulation}
%-------------------------------------------------------------------------------

This section presents the manipulation landscape encompassing all three elements (cf. \S\ref{sec:intro}) of SMM. While these elements are broadly defined in the literature, their relationships have not been thoroughly explored. We aim to shed light on the manipulation landscape and its intricate relationships. 
In Figure~\ref{fig:informationdisorder}, we present a visual representation of the manipulation landscape. Within this landscape, we depict false information utilizing \textit{information reporting, intent, event, content-context}, and \textit{claim type} represent false information (shown in green). Bots are represented by \textit{bot account type} and \textit{bot behavior} (depicted in light red). Additionally, we illustrate malicious campaigns through \textit{campaign controller (single bot or group-based)} and \textit{campaign promotion}. Note, campaign controllers employ various types of accounts and dissemination techniques to create and propagate false information content, thereby influencing user beliefs.

To thoroughly understand the landscape, we first provide definitions for the \textit{types of OSN users}, \textit{OSN account types}, and the \textit{psychological beliefs}. By doing so, we contribute to the existing knowledge in this area by presenting a comprehensive view of the manipulation landscape and highlighting the relationships between its constituent elements.
We then provide detailed definitions of each SMM-related element in the subsequent sub-sections.

\paragraph{Types of OSN Users.} 
Users on OSNs exhibit their unique persona~\cite{jarrahi2022evaluating}: \textbf{(i) Naive user.} A user that consumes information on OSN and often spreads unverified information with connections without knowing the authenticity. They typically hold strong psychological beliefs with which they react to the post and are referred to as a \textit{receptive audience} to falsity. \textbf{(ii) Conscious user.} This category of OSN users consumes information but tries to verify the authenticity and avoids sharing with connections unless they are sure of the veracity. \textbf{(iii) Malicious user.} A malicious user is the false content producer on OSN, and its success depends on manipulating a naive user successfully into believing false information. 

\paragraph{OSN users by Account Type.} OSN users, as described above, manage their accounts in various ways:
\label{subsec:osnusers}
\textbf{(i) Human accounts.} Human user accounts are genuine accounts entirely run by humans to interact on OSN.
\textbf{(ii) Bot accounts.} 
Bots are fully automated accounts that are fake and used for specific tasks. Currently, bots on OSN, such as Twitter, range between 5\%\cite{us_sec_twitter} to 25\%\cite{keller2019social}. (See Appendix~\ref{appendix:bot_twitter} for details).   
\textbf{(iii) Hybrid or Cyborg accounts.} Partially automated like cyborg accounts and called \textit{human-assisted bots} or \textit{bot-assisted human} accounts \cite{chu2012detecting}. For instance, a celebrity or media person account may use software schedulers to post at a particular time and event on their behalf. 

\begin{table*}[h]
\caption{First Element of OSN: False Information}
\centering
\scalebox{0.85} {
    \begin{tabular}{p{0.005\linewidth} p{0.024\linewidth} p{0.19\linewidth} p{0.87\linewidth}}
  \hline
  \hline
    &\multicolumn{2}{l}{\textbf{SMM Elements}} & \textbf{Definition}\\ \hline
    \hline

    1.1.&& \textbf{Information Reporting}  & Information reporting through journalists and experts involves multi-layered information collection, curation, and transformation that often induces elements of human-led mistakes, such as:\\

    & 1.1.1.& Error Reporting & Unintentional display of inaccurate numbers or statements about an event or an individual that might lead to a negative impact~ \cite{cnn_error}. \\
    
    & 1.1.2.& Missing Report. & Perspective is partially explained that leads users to make wrong decisions~\cite{missing_information}. \\
    
    & 1.1.3. & Bias Reporting & This involves bias for one party rather than presenting a neutral face causing echo chambers~\cite{bias_information}. \\
    
    1.2. && \textbf{Event-type} & Information covering a topic or an event, and there exist three high-level event types, such as: \\
   
    & 1.2.1. & Predictive/ Emerging & In this element, adversaries aim to utilize vulnerable topics such as elections or health crises to maximize the spreading of false information. \\
    
    & 1.2.2. & Random & This involves predicting the topics of stories that re-surface multiple times on OSN. \\
    
    & 1.2.3. & Time-critical & Some false information-related events are time bound for example, when bad actors capitalize on the situation by duping relief funds~\cite{turkey}. \\
    
    1.3. && \textbf{Spread-intent} & Multiple intents fabricate the stories propagated on OSN, e.g., a user can generate false information intentionally or unintentionally ~\cite{ireton2018journalism}. \\

    &1.3.1 & Misinformation & A piece of untrue statement or false information shared with no intention to cause harm. In this case, most users are unaware of information falsity and assume it to be accurate due to confirmation bias and naive realism. \\
   
    &1.3.2 &Disinformation & Disinformation is false information shared with a deliberate intention to cause harm to OSN. In terms of severity, disinformation has a devastating impact, e.g., COVID-19 disinformation. \\

    &1.3.3 &Malinformation & It involves using true information in a harmful way to attack an individual's reputation, e.g., Wikileaks released Hillary Clinton's emails that brought several ramifications from bad actors~\cite{hillary}. \\
   
    1.4. && \textbf{Claim-types} & False information claim covers a broad spectrum, i.e., from rumors to fake news.
    Over the years, multiple categories of types that differ in claim nature have surfaced, such as:\\ 
    
    &1.4.1 &Rumor & Rumors are unsubstantiated information \cite{starbird2014rumors}. \\

    &1.4.2 &Hoax & It delivers a form of trickery using an artifact that aims to convince users that a piece of false information is authentic.\\

    &1.4.3 &Urban legend& These are short narratives explaining unusual events that happened to an individual in the past representing the fears of the population. \\

    &1.4.4 &Spam& Spam is unsolicited messages sent on OSN, typically to many users, for advertising, phishing, or spreading malware. \\
    &1.4.5 &Myth& Myth is a form of supporting a false belief over the years without support or evidence to prove it. \\
    &1.4.6 &Propaganda& It frequently involves manipulating social motives in OSN by shaping the content to influence users' attitudes, values, and knowledge. \\
    &1.4.7 &Fake opinions& It relates to fake reviews that defame or degrade an entity's status.  \\

    &1.4.8 &Conspiracy& It is a secret action of multiple OSN users taken together to achieve a malicious goal. \\
    
    &1.4.9 &Fake News& Fake news is a proven untrue statement that many believe to be trustworthy. \\
    1.5. && \textbf{Content-context} & False information utilizes context for its viral propagation and to convince the OSN users~\cite{firstdraft}. \\
    
    &1.5.1 &Satire and Parody& It is an ironic message shared on OSN. It holds no intention to harm but still has the potential to fool OSN users. \\
    
    &1.5.2 &False Connection&
    When headlines, visuals, or captions do not support the content. These tactics and techniques (such as clickbait) amplify and increase the impression reach of the post or the story on OSN. \\
    
    &1.5.3 &Misleading content&
    An inaccurate presentation of information to favor an issue or an individual, e.g., wrong statistics showing high differences over small values to favor the agenda and mislead the user. \\
    
    &1.5.4 &False context&
    It uses true information in a twisted form to represent a false narrative that is likely to be believable, e.g., using wrong captions on another image that replaces the original context. \\
     
    &1.5.5 &Imposter content&
    Imposter content is when cloned profiles of genuine sources, such as news outlets, are used. The intuition behind this is to gain an audience base already familiar with the news outlets. \\ 
    
    &1.5.6 &Manipulated content&
    Manipulated content is photoshopped images, doctored videos, or AI-generated realistic fake content. For a naive user, it is a challenge to notice manipulations in real time. \\
    
    &1.5.7 &Fabricated content& Adversaries spend resources and time creating false stories or completely fake articles, mainly in the interest of causing harm. It involves the usage of inauthentic and low-credibility content. \\ \hline

  \hline
  \end{tabular}
  }
  \label{tab:osnlandscape1}
  \vspace{-1em}
  \end{table*}

\vspace{-2em}
\paragraph{OSN user psychological beliefs.} 
Researchers have invested considerable time in understanding user psychology concerning fake news spread as various psychological beliefs of users trigger this participation~\cite{confirmation, naive, homophily, echochamber}. These include: 
\textbf{(i) Confirmation bias.} A phenomenon in which a user believes a claim or information that confirms their presumption about the claimed entity \cite{confirmation}. For instance, users are most likely to believe that a powerful entity or an organization created the COVID-19 virus if it aligns with their existing belief. 
\textbf{(ii) Naïve Realism.} It is a state where users believe only their assumptions are the fact and bases of reality while differing from the opposing views \cite{naive}.
\textbf{(iii) Homeostasis.} Homeostasis is acting like other connections (surrounding friends or society) to maintain balance and stability—generally, gullible users in OSN tend to agree with their network.
\textbf{(iv) Homophily.} A phenomenon in which similar people interact with each other and disassociate with dissimilar people \cite{homophily}. This phenomenon can lead to polarized groups causing echo chambers \cite{echochamber} or \textit{filter bubbles} that compel OSN algorithms to promote similar content. 

\begin{table*}[]
\caption{Second Element of OSN: Bots Usage on OSN.}
\centering
\scalebox{0.85} {
    \begin{tabular}{p{0.005\linewidth} p{0.024\linewidth} p{0.14\linewidth} p{0.91\linewidth}}
  \hline
  \hline
    &\multicolumn{2}{l}{\textbf{SMM Elements}} & \textbf{Definition}\\ \hline
    \hline
    2.1. & & \textbf{Bot-types}  & Automated accounts or bots can serve multiple purposes, each built with a precisely defined goal, from a simple spamming bot to a social bot that interacts with OSN users like an actual human~\cite{botornot,botsurveypaper}. Bots can amplify information (or false information)~\cite{shao2018spread}. The following bot types are:\\
     
    &2.1.1.& Spam bots& Tasked to share malicious links, malware, false information, or hijack OSN trending topics.\\
    &2.1.2.& Follower bot& It is tasked to increase the social status of another account or its credibility.\\
    &2.1.3.& Astroturf bot& Accounts that support a particular candidate or event and create artificial support to look like it originates from the grassroots. \\
    &2.1.4.& Clone bots & Also known as doppelganger bots, clone profiles of ordinary users for nefarious reasons.\\
    &2.1.5.& Political bots& Tasked to interact in political discussions and manipulate public opinions by pushing propaganda.\\
    &2.1.6.& Sybil bot& One single entity creates and operates many accounts on OSN.\\
    &2.1.7.& Financial bots& It is tasked to spread economic disinformation and try to fluctuate trading markets.\\
    &2.1.8.& Scam Bot& It is tasked to lure na\"ive users into romance phishing to extort money or ultimately hijack accounts.\\
    &2.1.9.& Social Bot& It is a software-based account that mimics users' actual behavior and manipulates public opinion. \\
    &2.1.10.& Trigger Bot& A new family of spambots that only gets triggered on certain keywords on OSN, such as the keyword `metamask' or `gfx' in Twitter. It is tasked to lure na\"ive users into exposing their cryptocurrency wallet keyphrase.\\
    2.2 & & \textbf{Bot Behavior} &Different Bots may behave differently~\cite{cresci2023demystifying}. For example, a benign bot would notify genuine information time-to-time. However, forms of bot behavior often used by malicious users are: \\
    &2.2.1.& Focused/ Exploitative& As Khaund mentioned~\cite{9518390}, bots dedicatedly perform similar actions and post on a focused theme or topic. These bots are aggressive in their behavior, which is often easily measured. \\
    &2.2.2.& Mixed/ Explorative& Sophisticated bots~\cite{akhtar2022machine} post on several topics to remain undetected. It is complex to identify the real ideological nature of the profile. \\
   
  \hline
  \hline
  \end{tabular}
  }
  % }}
  \label{tab:osnlandscape2}
  % \vspace{-em}
\end{table*}

\begin{table*}[h]
\caption{Third Element of OSN: Malicious Campaigns.}
\centering
\scalebox{0.85} {
    \begin{tabular}{p{0.005\linewidth} p{0.024\linewidth} p{0.14\linewidth} p{0.91\linewidth}}
  \hline
  \hline
    &\multicolumn{2}{l}{\textbf{SMM Elements}} & \textbf{Definition}\\ \hline
    \hline
    3.1. & & \textbf{Campaign Controller} & False information is often fueled by fake accounts that control false information campaigns~\cite{bovet2019influence}. With the help of computational tools and resources, they act in the following ways:\\ 
    &3.1.1.& Single Bot/ Core Bot& This fake account acts individually or as a botmaster that controls the false information campaign. They are the mastermind behind the production and spread of false information on OSN. \\
    &3.1.2.& Group-based/ Peripheral& These act as proxy bot accounts to increase the influence of the core bot on OSN. They act as slaves to the botmaster, mainly disseminating or amplifying false information and the core bot profile. \\
    3.2. & & \textbf{Campaign Promotion} & In general, campaigns lead to information propagation to a broader audience. Therefore, many strategies undergo for having a successful campaign~\cite{akhtar2022machine}. The two common broad approaches are:\\
    &3.2.1.& Organic& 
    The organic campaign stems slowly from OSN and takes a long period to reach a wider audience. It builds a correlation with OSN algorithms to gain attention from followers. It does not employ cost.\\
    &3.2.2.& Inorganic& This campaign uses paid promotions and reaches to wider OSN population quickly. Thus, it gains more reactions and interaction with the users (including naive and conscious users).  \\
  \hline
  \hline
  \end{tabular}
  }
  \label{tab:osnlandscape3}
  \vspace{-0.5em}
\end{table*}

\paragraph{OSN Manipulative Actors.} 
{OSN consists of different OSN users, users by account type, and users with various psychological beliefs. This is the OSN vector that manipulative actors manipulate. A manipulative actor could have properties belonging to single or multiple attributes, such as a manipulative actor could be human, operating a bot account in a coordinated fashion with other state-sponsored accounts. In brief, these attributes are: 
    \textbf{(i) Malicious vs. non-malicious users.}
    Prior research has identified that users who play their role in disseminating false information are motivated by their \textit{intent}. However, users can forward false information, irrespective of their intent. Starbird et al. \cite{starbird2019disinformation_trolls} have referred to the non-malicious category of users (such as naive users) as `unwitting human collaborators of false information.'
    \textbf{(ii) Bots vs. normal human accounts}
    Accounts on OSN operate with differing levels of \textit{usage}; one such can often use automated tools to scale up the activity. Moreover, we acknowledge that manipulative actors need not certainly use automation and can be simple normal human accounts \cite{starbird2019disinformation_trolls}. 
    \textbf{(iii) Coordinated behavior vs. individual activity.} 
    Manipulative actors on OSN could team up together to form a controlled cluster by exhibiting orders in a coordinated behavior \cite{starbird2019disinformation}, or they can act as individuals towards their goals. The coordinated behavior of manipulative actors is generally more seen from the \textit{behavior} of malicious users. In other words, non-malicious actors do not follow any motive to act in coordination. 
    \textbf{(iv) Foreign state-sponsored actors vs. individual.} 
    Often, the manipulative actors are driven by either personal agenda or motive to conduct their activity on OSN, or other times; foreign states or organizations fund them \cite{zannettou2019let} in terms of financial assistance to set up accounts or purchase several fake accounts and APIs, often referred as advanced persistent manipulators \cite{franccois2020actors}. The US government has called these accounts `weapons of mass distraction' \cite{nemr2019weapons}.  

Next, we define elements of false information, namely: information reporting, intent, event, content context, and claim type (as shown in Table~\ref{tab:osnlandscape1}) that aid actors in manipulation. One such instance of 
false information may include various content types, such as manipulated content, misleading context, or satire, to create propaganda, fake news, or rumors. Note that all categories of information that deceive users are deceptive and harmful content \cite{franccois2020actors}. For the same reason, we did not opt out of categories that have truth values but still tend to misinform, fool, or deceive users, such as satire or parody.

With the terms defined in Table~\ref{tab:osnlandscape1}, false information binds all elements together and poses a concern to OSN. As mentioned, manipulative actors could use automated accounts (bots) or genuine accounts. However, in this research, we only study the prevalence of bots even though human accounts also propagate false information. The reasoning is to share insights on the illicit use of computational means that aid viral deception, as evidenced in prior research~\cite{franccois2020actors,ferrara2016rise, akhtar2022machine, himelein2021bots, shao2018spread, bovet2019influence}. On the other hand, several canonical papers exist, delving into the psychological factors behind human accounts spreading false information~\cite{glenski2018propagation, 10.1371/journal.pone.0239666, o2019misinformation, Starbird_Arif_Wilson_Van_Koevering_Yefimova_Scarnecchia_2018, zannettou2019let}. Thus, the following Table~\ref{tab:osnlandscape2} defines the second element of SMM, i.e., bots. Malicious actors can create bots for various tasks such as spam, financial fraud, or to spread disinformation.} These account types can behave either focused or explorative as the manipulator commands. 

As shown in Figure \ref{fig:informationdisorder}, bot account and behavior typically demonstrate the computational medium the manipulator uses to manipulate OSN. On the other hand, malicious campaigns are the driving force behind various manipulations. Bot controllers can be individuals or groups focused on spreading false information. Moreover, they take the help of various promotion techniques on OSN, such as organic or inorganic promotion, to artificially support their agenda. The elements of malicious campaigns are presented in Table~\ref{tab:osnlandscape3}. (See Appendix~\ref{appendix:osnlandscape} for works in respective landscape)

All the points in the tables highlight the different components of manipulations and paint a picture of the whole SMM landscape. We believe that using our systematization, (i) scholars can incorporate diverse viewpoints from existing interdisciplinary research in their solutions to mitigate SMMs, and (ii) OSN providers can develop new emergent tools, informed strategies, and novel policies toward a safe OSN. Due to the intricate interplay among users, psychology, and technology, research in this field is not confined to a single discipline---instead extends to interdisciplinary and transdisciplinary approaches~\cite{carley2020social}. The next section explores false information detection methods followed by the bot and malicious campaign detection systems.
% \vspace{-1em}
\section{First element of SMM: False Information Spread in OSN}
\label{sec:false_information}
The problem space we discuss involves a complex relationship between psychology, communication, and human-machine interaction. This space naturally extends to the concept of ``six degrees of separation''  proposed by Frigyes Karinthy in the early 20th century. This concept states that any two people on Earth are six or fewer connections away from each other \cite{travers1977experimental, sixdegree}. Later, Cheng \cite{fivedegree} identified the same phenomenon on Twitter and discovered that any two online users are five connections away from each other. We anticipate that this observation benefits bad actors seeking to spread false information quickly on the OSN. As expected, false information spreads rapidly and deeply on OSNs. So far, many efforts have been proposed to tackle the problem of false information. Given that detection of false information requires an interdisciplinary approach---we discuss the approaches that employ research from various fields, such as information theory, crowd intelligence, and machine learning. These works share the same goal but differ significantly in metrics and methodology. Thus, in this section, we aim to answer the research question  \textbf{RQ1}: \textit{What are the characteristics of the spread of false information?} and \textbf{RQ2}: \textit{What are the most commonly used effective techniques and features for detecting false information?}
In response to the first question, we reviewed papers on understanding the spread of false information (diffusion) and its properties (dynamics).
In response to the second question, we follow taxonomy from simple to increasingly complex approaches to false information detection.

\vspace{-1em}
\subsection{False Information Detection Approaches}

\subsubsection{Diffusion Dynamics of False Information}
\label{subsubsec:ddfi}

In information theory systems or social network analysis, the examination of \textit{diffusion} is the most effective method for measuring the influence of information in a network. More
precisely, the apparent term is diffusion dynamics, which
refers to studying information spread (cascades) and its properties \cite{doi:10.1126/science.aap9559}. Ideally, a single cascade (a story tweeted independently) motivates researchers to study diffusion. For example, the Boston Marathon bombing event \cite{starbird2014rumors} or the infamous ``Pizzagate'' conspiracy theory \cite{pyrhonen2020conspiracies} motivated researchers to infer conspirator evolution strategies behind the stated events in OSN. While we notice unique insights from individual topic-based diffusion studies, {but there is a} lack of global observations within various false information stories. 

Thus, Vosoughi et al. \cite{doi:10.1126/science.aap9559} investigated the propagation of false news (rumors) on Twitter among the 126,000 stories between 2006 and 2017. 
The authors' evaluation of diffusion dynamics of both categories of rumors reported that false stories ``\textit{tend to spread faster, more broadly, deeper and reach more people}'' than true stories \cite{doi:10.1126/science.aap9559}. They found that users who spread false information are newer accounts, less active on Twitter, have less following/follower (FF) ratio, and are less likely to be verified. 
To understand the reasoning behind retweets of falsity, the authors assessed the novelty of information as it attracts human attention. The result suggests that ``\textit{falsity was more novel (high uniqueness) than true news}.''

\paragraph{Properties of diffusion dynamics.} We need methods to understand diffusion's properties and quantify the influence of false information. For this purpose, we extract information on the diffusion properties driven by technical papers. Kai et al.~\cite{fakenewsnet} conducted one such work. The author published a tweet dataset for fake news detection that includes user reactions with temporal information. This dataset is an excellent resource for understanding the structural properties of the spread. Therefore, we selected some particular topic-based false and real news stories, such as politics, LGBTQ+, religion, and sports, from the Kai et al. \textit{in situ} dataset to reproduce temporal graphs of retweets, tweets, and replies. Figure~\ref{app:2} displays the tweets, retweets, and replies count. We agree with the conclusion of Kai et al., who observed that \textit{``fake news attract a more sudden increase in retweets and fewer replies compared to real news, which gains a steady increase in the number of
retweets''} as shown in Appendix~\ref{app:information_diffusion} in Figures~\ref{app:real} and \ref{app:fake}. In other words, real news has a more uniform distribution throughout the life cycle of news distribution in OSN compared to fake news, as shown in Figures~\ref{app:real1} and \ref{app:fake2}. Other scholars such as Cheng et al. \cite{cheng2021causal} attribute the sudden burst of tweets or retweets to user's behavior and activities, acting as \textit{confounders}. The Confounder is a hidden causal variable responsible for the cause (such as the user's susceptibility to fake news). Cheng et al. found that \textit{verified, status count, friends count, and user organization information are essential causal spread properties.} 

\paragraph{Identifying actors in diffusion dynamics.} 
To identify users involved in false information spread on OSN, we categorized works that examined user profile characteristics and susceptibility levels to determine who shares false information and why \cite{avvpyscholinguistic1,avvpsycholinguistic2}. Kai et al. \cite{24shu} analyzed explicit features (profile-related, content-related, and network-related) and implicit features (gender, age, and personality traits) to explore the correlation between user profiles and fake news. Their findings revealed that users who registered earlier were more prone to trust fake news, while newer accounts tended to spread more real news. Additionally, users who trusted more real news were more likely to be active and expressive, while older and female users were likelier to trust fake news. The study also found that users who shared fake news had a higher agreeability based on the five-factor model, while users who shared real news showed extraversion and openness.

On the other hand, Shen et al. \cite{susceptibility} established a framework to determine the susceptibility levels of users in response to fake news on OSN. The result indicates that the high susceptibility level of the center nodes (influential nodes) was highly correlated with the susceptibility level of the entire network. This means the network remains less susceptible if the influential node avoids sharing false news. This set of observations addresses the research question of which users are more likely to trust/distrust fake news. 

\begin{tcolorbox}[colback=blue!9!white, top=0pt, bottom=0pt, left=0pt, right=0pt]
\textbf{Takeaway 1.} \textit{User's demographics, personality traits, susceptibility level, and user's position in the network play a significant role in identifying malicious actors in spreading false information.}
\end{tcolorbox}

\subsubsection{Knowledge-based False Information Detection}
We are aware that false information lacks verification. In pursuit of fact-checking, various mechanisms have been proposed, including knowledge-based approaches, that verify the knowledge in the to-be-verified content with fact-checked knowledge.
This endeavor is bolstered by two knowledge-based strategies for detecting false information, one involving manual fact-checking and the other using crowdsourcing methods, with differing levels of proficiency and scope.

\paragraph{Manual-based expert fact-checking.} The most effective way to detect and prevent misinformation is for experts to debunk false stories in real time. Top fact-checking organizations, such as PolitiFact~\cite{PolitiFact}, employ a rigorous process that considers multiple details before reaching a verdict. They scrutinize source artifacts to verify their authenticity. However, this approach to fact-checking has several limitations: 
% \begin{description}

(i) \textit{Manual fact-checking is time-consuming and labour-intensive.} The volume of information on OSNs and the reactive nature of manual debunking present challenges for timely fact-checking and proactive prevention of misinformation.

(ii) \textit{Experts must maintain transparency and impartiality, without biases} to maintain their credibility and trustworthiness. They must use reliable sources and transparent and replicable methods for debunking false information.

(iii) \textit{Identifying whether a trend in OSN is organic or maliciously promoted is daunting.} 
    Manual analysis of coordinated activity towards false information is infeasible in real time, necessitating automatic tools or crowd wisdom for fact-checking scalability.

\paragraph{Crowd-based fact-checking.} 
Crowd-based fact-checking is another approach that gathers many regular individuals to assess the content and provide their verdict. The intuition is that while regular individuals may not be proficient at distinguishing between true and false news, they can collectively identify manipulations in real time, making them valuable in fact-checking. Such collective intelligence allows for scalable fact-checking that complements the work of professional fact-checkers, as evidenced by recent research \cite{doi:10.1126/sciadv.abf4393}. We see the usage of a crowd in one of the widely popular fake news datasets, \textit{CREDBANK}---that employed 40 Turkers (workers from Amazon Mechanical Turks) for annotation assessment \cite{mitra2015credbank}.

However, Saeed et al. \cite{saeed2022crowdsourced} recently argued that crowd intelligence (on platforms such as Birdwatch) does not guarantee consistency but only helps provide additional sources of evidence for professional fact-checkers. Similarly, Stein et al. \cite{stein2023realtime} shared the limitation of using crowd on Birdwatch, as users may propagate their ideology and falsely support false information to alter the fake news veracity. It is striking that only a few research have examined the effectiveness of crowd-based fact-checking. 

\vspace{-0.5em}
\subsubsection{Content and Social Context-based}
Our survey of academic papers found that the automatic detection of fake news depends primarily on the content of false information. In all these cases, studies resort to feature engineering on various aspects of the content, such as writing style, readability, sentiment, source credibility, and the auxiliary social context derived from the fake content \cite{10.1145/3393880}.

\paragraph{Writing style, language, and linguistics.}
The language used in OSN posts or news articles can provide information about the text's credibility. Fake news often uses hyperbolic language, emotional appeals, and misleading information, while legitimate news uses neutral and fact-based language. In particular, previous research has transformed text content into vectors or embeddings using various models for classification purposes. For example, Akhtar et al. \cite{akhtar2022machine} used (TF-IDF) and (BERT-based) sentence embeddings to transform the tweet's text into contextual vectors. Whereas Sajjad et al. \cite{rezaei2022early} considered a mixture of feature generation efforts such as using topic modeling, calculating similarity, and adding the count of words, sentences, and sentiment score of the overall text fed into an ensemble-based model. Though promising results, adding other social indicators in the classification may improve accuracy. Such social indicators are social context information that utilizes network features, including the degree of in / out, page rank, and early user reactions, as Nguyen et al. \cite{tuan2020fakenews} extracted network features and found that combining these features provides better results than using only text-based classification to detect fake news.

\begin{tcolorbox}[colback=blue!9!white, top=0pt, bottom=0pt, left=0pt, right=0pt]
\textbf{Takeaway 2.} \textit{Works identifying false information using content has leveraged contextual text transformations, semantic analysis, and insights from social context for detection.}
\end{tcolorbox}

\vspace{-0.5em}
\paragraph{Readability.} A news article's readability can indicate its complexity level and target audience. Fake news articles often use straightforward language to reach a broad audience, whereas legitimate news articles use more complex language and may target a more specific audience. Qayyum et al. \cite{qayyum2022deep} reported low readability scores for focused toxic profiles (exploitive user) using semantic and grammatical correctness of toxic posts. We conjecture that their findings give the reasoning for Kai et al.\cite{fakenewsnet} result that reported fake news attracts many retweets since producing new fake content is more challenging than actual news. Malicious actors would be aware of such language-based detection and may realize adversarial training on their content to pass language scores as a bare minimum requirement. However, we have yet to learn any comprehensive work focusing on all or many strategies that bad actors employ. 

\begin{tcolorbox}[colback=red!9!white, top=0pt, bottom=0pt, left=0pt, right=0pt]
\textbf{Research Gap 1.} \textit{The current research is lacking information regarding the strategies that bad actors employ to improve the readability of false information in the age of AI-generated content~\cite{Ferrara_2023}.} 
\end{tcolorbox}

\vspace{-0.5em}
\paragraph{Sentiment, Bias, and Stance.} Fake news often expresses a biased or emotional sentiment, while legitimate news articles express a neutral or balanced sentiment \cite{ZHONG2023102626}. Similarly, as seen in crowd-based detection, initial crowd reactions offer insight into false information detection. At the same time, earlier studies show that a user's belief, such as confirmation bias, leads to deciding the story's verdict instead of others' influence. However, in recent work, initial reactions (sentiment and posture) of users exposed to the information often lead to the decision-making of subsequent users \cite{stein2023realtime}. Furthermore, we observe that using the originator's bias improves the detection \cite{beyondnewscontent}. However, relying solely on sentiment or stance is insufficient to combat misinformation effectively. The originator and the spreader of misinformation are potential clues for detection.

\paragraph{Source Credibility.} We do have some information that the source of misinformation is vital in determining its credibility. Using the profile feature, such as the account lifetime, the number of followers (to measure their influence), and the number of friends, which measures their sociality behavior, interaction activity, and political inclination, Jarrahi et al. \cite{jarrahi2022evaluating} demonstrate that the false news publisher's feature improves overall false information detection. However, an artificially produced profile with fake followers and friends can impact source credibility algorithms. Most of the studies do not consider the influence of fake or bot-controlled accounts on disseminating fake news \cite{akhtar2022machine}. However, we agree that sentiment and source features acting as social context improve false information detection.

\begin{tcolorbox}[colback=blue!9!white, top=0pt, bottom=0pt, left=0pt, right=0pt]
\textbf{Takeaway 3.} \textit{Social context adds complementary information to content-only false information detection by considering the broader environment in which the information is shared or consumed.}
\end{tcolorbox}

\subsection{Discussion}
\label{subsec:false_discussion}
In summary, 
we realized that social-context information helps detect false information. Users' reactions can act as a social sensor---different responses to different stories \cite{rizoiu2018sir}. However, waiting for more user reactions (or interactions) could delay early detection, which limits effective mitigation of the spread of misinformation. Therefore, most attempts are made to detect false information using \textit{content} for early detection. On the other hand, we know from research that bots and fake accounts catalyze the spread of false information. Kai et al. \cite{fakenewsnet} found 22\% of accounts identified as bots responsible for disseminating fake news in their dataset. Moreover, we believe that bots or fake accounts could generate social context, similar to genuine users. Therefore, bots should be detected before the spread of misinformation.

\begin{tcolorbox}[colback=red!9!white, top=0pt, bottom=0pt, left=0pt, right=0pt]
\textbf{Research Gap 2.} Social context can be helpful, but current research has scarce literature on false social context generated by bots. Bots can amplify information and make a network of inauthentic accounts~\cite{boichak2021not}.
\end{tcolorbox}

To this end, Table \ref{tab:lit_works} analyzes works in false information detection 
(see the characterization of literature method details in Appendix~\ref{appendix:characterization}).
From Table \ref{tab:lit_works}, we conjecture that most work focuses on textual-based feature extractors such as n-grams, (TF-IDF), and word embeddings like (BERT), whereas (Resnet-50), (VGG-16), (VGG-19) have been widely used as feature extraction from images. However, most of the research in this area has focused on Twitter rather than other OSNs, possibly due to the availability of open-access data from Twitter and the platform's continued importance as a source of information. Furthermore, we note that no single study focuses on \textit{Tier 3} data collection (extracting account's neighbors' information), possibly due to API rate limit issues. Even though exemplary work exists in false information detection, the focus primarily seems to be detecting the ``content of news articles'' (of the mainstream media) rather than social media data. Another limitation observed is that most work is built on fact-checking content from American-based fact-checking organizations like PolitiFact~\cite{PolitiFact} or Snopes~\cite{Snopes}, favoring English content verifiability rather than a generic model for curbing false information. Hence, a language-agnostic model is required for false information detection.

\begin{center}
\begin{table*}[h]
    \centering
\caption{\small Analysis of state-of-the-art and repository in False information. \textbf{Data Collection (DC)}: \fullcirc[0.8ex] = Manual Collection; \emptycirc[0.8ex] = Secondary Data; \halfcirc[0.8ex] = Both used. \textbf{Data Collection Tier}: \emptycirc[0.7ex] \emptycirc[0.7ex] \emptycirc[0.7ex] = Tier 0; \fullcirc[0.8ex] \emptycirc[0.7ex] \emptycirc[0.7ex] = Tier 1; \fullcirc[0.8ex] \fullcirc[0.8ex] \emptycirc[0.7ex] = Tier 2; \fullcirc[0.8ex] \fullcirc[0.8ex] \fullcirc[0.8ex] = Tier 3. \textbf{Availability}: \fullcirc[0.8ex] = Data Available; \ding{73} = Code Available; \ding{74} = Both Available. \textbf{FS + Model} = Feature Selection + Model type. 
\textbf{Modality}: \ding{46} = Text-based; \ding{41} = Image-based; \faCamera\ = Multimodal data (Text+Image). \textbf{N/A} = Not Available. \ding{72} = Github Repository Star.}
  \centering
  \scalebox{0.72} {
  \begin{tabular}{c c c c c c c c c c c c c c c c}
  \hline
  \hline
    \mcrot{1}{l}{60}{Work} &	\mcrot{1}{l}{60}{DC*}	& \mcrot{1}{l}{60}{Platform} &	\mcrot{1}{l}{60}{DC Tier} &	\mcrot{1}{l}{60}{\#Size} &	\mcrot{1}{l}{60}{Classes} & \mcrot{1}{l}{60}{Availability} &	\mcrot{1}{l}{60}{FS + Model} &	\mcrot{1}{l}{60}{Popularity} & \mcrot{1}{l}{60}{Duration} & \mcrot{1}{l}{60}{Domain} & \mcrot{1}{l}{60}{Language} & \mcrot{1}{l}{60}{Modality}\\ \hline
    \hline
    
    \cite{fakenewsnet} & \fullcirc[0.8ex] & \shortstack{Mainstream+Twitter} & \fullcirc[0.5ex] \fullcirc[0.5ex] \emptycirc[0.4ex] & 432-16817 & \ding{173} & \ding{73} & Autoencoder + LSTM & 735 \ding{72} & N/A & \shortstack{Political, Entertain.} & English & \faCamera\ \\ \hline
    
    \cite{FND} & \emptycirc[0.7ex] & Mainstream & \fullcirc[0.5ex] \emptycirc[0.4ex] \emptycirc[0.4ex] & 12800 & \ding{173} & \ding{74} & \shortstack{bag-of-words, n-grams \& TF-IDF} & 293 \ding{72} & 2007-2016 & Variety & English & \ding{46} \\ \hline
    
    \cite{EXIF} & \halfcirc[0.7ex] & N/A & \fullcirc[0.5ex] \emptycirc[0.4ex] \emptycirc[0.4ex] & \shortstack{Random sample from\\ 400000 images} & \ding{173} & \ding{73} & \shortstack{EXIF-consistency 
Siamese network\\ + Resnet-50 then MLP} & 169 \ding{72} & N/A & N/A & N/A & \ding{41} \\ \hline

    \cite{UPFD} & \emptycirc[0.7ex] & Twitter & \fullcirc[0.5ex] \fullcirc[0.5ex]  \emptycirc[0.4ex] & 157-2732 & \ding{173} & \ding{74} & Spacy \& BERT using GNN & 177 \ding{72} & N/A & \shortstack{Political, Entertain.} & English & \ding{46} \\ \hline
    
    \cite{EANN} & \emptycirc[0.7ex] & \shortstack{Twitter \\ \& Weibo} & \emptycirc[0.4ex] \emptycirc[0.4ex] \emptycirc[0.4ex] & 514-9528 & \ding{173} & \ding{74} & \shortstack{Word embedd. + VGG-19} & 146 \ding{72} & \shortstack{May 2012\\-Jan 2016} & N/A & \shortstack{Chinese \\ \& English} & \faCamera\ \\ \hline
    
    \cite{areyoufakenews} & \fullcirc[0.8ex] & Mainstream & \emptycirc[0.4ex] \emptycirc[0.4ex] \emptycirc[0.4ex] & 45000 & \ding{178} & \ding{73} & TF-IDF +NLP+CNN & 109 \ding{72} & N/A & N/A & English & \ding{46} \\ \hline
    
    \cite{FND2} & \emptycirc[0.8ex] & Mainstream & \fullcirc[0.5ex] \emptycirc[0.4ex] \emptycirc[0.4ex] & 11000 & \ding{173} & \ding{73} & \shortstack{PCFG-only, Bi-gram TF-IDF} & 101 \ding{72} & \shortstack{1 Sept-30 Sept\\2015} & N/A & English & \ding{46} \\ \hline
    
    \cite{bertLiarplus} & \emptycirc[0.8ex] & Mainstream & \fullcirc[0.5ex] \emptycirc[0.4ex] \emptycirc[0.4ex] & 12800 & \ding{173} & \ding{73} & \shortstack{BERT + Siamese network} & 95 \ding{72} & 2007-2016 & Variety & English & \ding{46} \\ \hline

     \cite{fakeddit} & \fullcirc[0.8ex] & Reddit & \fullcirc[0.5ex] \emptycirc[0.4ex] \emptycirc[0.4ex] & 1,063,106 & \ding{177} & \fullcirc[0.8ex] & BERT + Resnet-50 & 91 \ding{72} & N/A & Variety & English & \faCamera\ \\ \hline 
    
    \cite{newsaudit} & \fullcirc[0.8ex] & Mainstream & \emptycirc[0.4ex] \emptycirc[0.4ex] \emptycirc[0.4ex] & <40,000 & \ding{174} & \ding{74} & \shortstack{N-Gram-based} & 86 \ding{72} & N/A & Politics & \shortstack{Multi-lingual} & \ding{46} \\ \hline
    
    \cite{FNDDL} & \emptycirc[0.8ex] & Mainstream & \emptycirc[0.4ex] \emptycirc[0.4ex] \emptycirc[0.4ex] & 562 & \ding{173} & \ding{74} & Bi-GRU \& Attention & 82 \ding{72} & N/A & Variety & English & \ding{46} \\ \hline
    
    \cite{WeFend} & \fullcirc[0.8ex] & \shortstack{Mainstream\\WeChat}& \fullcirc[0.5ex] \emptycirc[0.4ex] \emptycirc[0.4ex] & 20728 & \ding{173} & \fullcirc[0.8ex] & \shortstack{RL+NN feature extractor} & 78 \ding{72} & N/A & N/A & \shortstack{Chinese\\English} & \ding{46} \\ \hline
    
    \cite{fakebustercode} & \emptycirc[0.8ex] & Mainstream& \fullcirc[0.5ex] \emptycirc[0.4ex] \emptycirc[0.4ex] & 20800 & \ding{173} & \ding{74} & LSTM & 75 \ding{72} & N/A & N/A & Chinese & \ding{46} \\ \hline
    
    \cite{MVAE} & \emptycirc[0.7ex] & \shortstack{Twitter \\ \& Weibo} & \emptycirc[0.4ex] \emptycirc[0.4ex] \emptycirc[0.4ex] & 17000 & \ding{173} & \ding{74} & \shortstack{Multimodal Variational Autoencoder} & 61 \ding{72} & \shortstack{May 2012\\-Jan 2016} & N/A & \shortstack{Chinese \\ \& English} & \faCamera\ \\ \hline
  \hline
  \hline
  \end{tabular}}
  \label{tab:lit_works}
  \end{table*}
\end{center}
\vspace{-1.5em}
\section{Second element of SMM: Bots Usage On OSNs}
\label{sec:bot_detection}

Fake accounts and bots have impacted most of the OSNs, such as Twitter. We use Twitter's example to illuminate bot usage as it provides APIs and datasets, which are significant sources of information for researchers and journalists.

Twitter has had bots for over ten years, and throughout this time, we have seen the emergence of four generations of bots. Until 2011, bots were used to manage default profiles. These accounts had little personal information and fewer connections. They were mainly used for spamming the network and were easily distinguishable by OSN users, even by a naive user. Then followed the second generation of bots, which added personalization to profiles, and appeared more reputed and credible by having more social connections. This was the first sign of bots' evolution. Since 2016, the rise of social bots (third-generation bots) has changed the Twittersphere \cite{ferrara2016rise}. These bots participate in manipulations and mimic OSN users to appear genuine. In essence, most of these accounts followed a day-and-night cycle to display human-like behavior.

Moreover, these bots diverted themselves into different goals and behavior, such as spam, phishing, amplification, or follower bots. We conjecture that third-generation bots specific behaviors and roles helped researchers to build filters and robust classifiers against them. On the contrary, since 2020, a new wave of bots, known as adversarial, has emerged.
They act as adversaries and test the model's (bot detectors) capabilities to evade detection. As such, there is a lack of tools that confidently detects adversarial bots. In other words, these bots carry out their malicious activities while remaining undetected. Due to technical limitations, such bots lie only in discussions~\cite{10.1145/3184558.3191610}. Therefore, detecting and mitigating bots from OSN has been a major open problem in OSN research.

\subsection{Bot Detection Approaches}
The works on bot detection can be divided into two distinct categories~\cite{perspective}: \textit{inferential} and \textit{descriptive}. The two categories have the same goals but employ different approaches. The works in the first approach focus on generalized heuristics or prominent bots' features used in the detection process. In contrast, the second approach leverage case-by-case manually driven observations to detect evidence of bots' manipulations. Both inferential and descriptive approaches use various features in their model. We find five dimensions of feature groups for every account: user profile-based, content-based, temporal-based, devices-based, and network-based features (for details, see Appendix~\ref{appendix:features}).

\subsubsection{Inferential Approach}
The inferential approach uses heuristics and rules to detect bots. A few of the simple rules to detect bots are based on tweet/retweet-frequency~\cite{howard2016bots}, use of alpha-numeric strings in user name \cite{beskow2019its, chu2012detecting}, mismatch of name and gender \cite{walt_eloff}, use of third party clients \cite{chu2012detecting}, posting a duplicate tweet or malicious links \cite{chu2012detecting}. Based on some of these rules, Pandu et al. \cite{indonesiaelections} manually label 4000 accounts as a bot or not. However, we estimate that the bot characteristics in Pandu et al. work is neither exhaustive nor has a shred of solid evidence for bot characterization. For instance, the authors have assumed that the user is a bot if the account's username contains numbers. Although, in practice, Twitter, by default, includes numbers if a username is not unique. This is a part of the automatic name generation process, and many genuine users ignore changing it. Thus, such rules are simple, and adversaries know these hand-crafted features \cite{feng2022twibot}. 

While we agree with the inferential model, emphasizing the granularity level in the inferential approach plays an important role. For example, bot detection can focus on individual accounts or focus on detecting coordinating groups of accounts. This choice depends on the structure and focus of the bot detection method. 

\paragraph{Individual-based bot detection.} This approach trains classifiers to detect individual bot accounts using a defined set of features, aiming for micro-level detection. However, Sayyadiharikandeh et al. \cite{sayyadiharikandeh2020detection} identified that bot datasets vary regarding class labels and characteristics. For example, spambot would have many tweets and retweets compared to a follower bot. Combining one common rule for all bots would compromise the precision and performance and increase the false positives \cite{cresci2023demystifying}. Thus, Sayyadiharikandeh et al. \cite{sayyadiharikandeh2020detection} introduced a novel method by utilizing ensembles of specialized classifiers. Specialized classifiers detect different bot classes using their respective essential features. Although this model outperformed the previous work \cite{featureanalysis}, it is unclear to view the model performance on a new class of bots. For the same reason, we believe retraining would be required on a specialized classifier for the new bot class.

In another study, Walt and Eloff \cite {walt_eloff} simulated fake accounts based on the mismatch between name and gender or falsely provided location data to build an accurate fake account detector. However, it is well known that malicious users scrap OSN public data to create genuine look-alike profiles \cite{HATFIELD2018102}; therefore, the work overestimates the features for accurately detecting fake human identities. Moreover, the works mentioned above do not focus on coordinating bots that may look normal under individual inspection. Thus, another approach focus on group-based bot detection.

\paragraph{Group-based bot detection.} It focuses on the macro level, which notices patterns among multiple individual accounts. 
As such, Shao et al. \cite{adverarial_cost_shao} use unsupervised techniques that include several steps---clustering of similar accounts and manual annotation of accounts for classification purposes—however, the chosen feature derived from profile metadata can easily be tampered with. Similarly, Cresci et al.\cite{common7, dna_modeling_cresci} proposed an innovative method that uses digital DNA sequences of the user's timeline data to detect coordinating bots. Though this work produces significant results, it only detects a special kind of bot, i.e., spambot. Any other bot that does not fill much timeline activity, such as a follower bot, cannot be determined by digital DNA \cite{cresci2023demystifying}. At the same time, the inferential approaches may also introduce false positives due to poor generalizability. Thus, \textit{descriptive} follows that provides true traces of manipulation of bots.

\begin{tcolorbox}[colback=red!9!white, top=0pt, bottom=0pt, left=0pt, right=0pt]
\textbf{Research Gap 3.}\textit{ There is a lack of literature on models that detect bots and considers bots diverse characteristics and labels.}
\end{tcolorbox}

\subsubsection{Descriptive Approach} 
\label{subsec:descriptive}
These approaches mainly use manual observations aided by clustering techniques or data analysis tools. As such, Graham et al. \cite{graham2020like} utilized manual observations to analyze campaigns while studying the ``\textit{coordinated spread of covid-related disinformation}''. They identified highly coordinated account clusters based on users' co-retweet frequency. On manual examination of multiple clusters, they found some clusters to be spam bots, self-identified bots, and Turkish disinformation bots. Compared with the inferential methods, a single descriptive work can find a variety of bots. However, this approach is not feasible in real-time at a large scale as it heavily depends on manual observations.

In a similar study, Arif et al. \cite{arif2018acting} built two clusters of accounts that supported either \#BlackLivesMatter or \#BlueLivesMatter. Within these contradictory clusters, they searched for Internet Research Agency (IRA) accounts (accounts that caused manipulations in the 2016 US Elections). Surprisingly, these accounts were at the core of both clusters. On manual analysis of these accounts, they found these accounts shape the online discussion by inciting online communities for violence or cultivating fake discussion between two IRA accounts to model anger and division \cite{arif2018acting}. In the subsequent work, Starbird et al. \cite{starbird2019disinformation} found strategic communities of bots, trolls, and unwitting human accounts, to be the significant collaborators of disinformation. These accounts often used propaganda websites to push their narrative or even attributed alternative narratives to sensitive events. 

However, descriptive approaches require analysis based on knowledge and the context of the chosen campaign category. Only then can a researcher develop evidence of manipulations that solidify their claim (see Appendix~\ref{appendix:case_study} for a case study on trigger bot, a new family of bots examined by our team). To this end, the discussion above highly depends on the feature engineering process. Not to mention collecting datasets from OSN faces API rate limit issues. Thus, different researchers focus on different features for detection. 

\begin{tcolorbox}[colback=red!9!white, top=0pt, bottom=0pt, left=0pt, right=0pt]
\textbf{Research Gap 4.}\textit{ There is a lack of methods that use both approaches for bot detection. Inference models can be fine-tuned by descriptive approaches to detect bots accurately..}
\end{tcolorbox}

\subsection{Discussion} 
This sub-section discusses a few observations on the bot's characteristics, the arms race between bot detectors and developers, bot dataset generation methods, critics, and comments on bot datasets and research.
\subsubsection{Generalizability Issues} 
OSN bots vary in behavior and final goals. Due to this, even robust models trained on one class of bots perform poorly on unseen bot accounts \cite{cresci2023demystifying}. This is further confirmed by a comprehensive generalizability study conducted by Echeverria et al. \cite{10.1145/3274694.3274738} on 9 datasets in a leave-one-botnet-out (LOBO) fashion. They found that models fail to detect unseen bot cases even though trained on all the other bots. The speculation is that the way these bot datasets are collected is inconclusive to say they represent all Twitter bots.  
Due to the high heterogeneity among bot classes, the generalizability of a model becomes a hindrance. Thus, new approaches that also focus on improving generalizability act as a good metric to assess the performance of the bot detector. 

\begin{tcolorbox}[colback=blue!9!white, top=0pt, bottom=0pt, left=0pt, right=0pt]
\textbf{Takeaway 4.}\textit{ New research should improve generalizability performance, as it is expected that the models should work on any OSN account rather than an account similar to the training dataset.}
\end{tcolorbox}

\subsubsection{Arms Race Between Bot Developers and Bot detectors}
\label{subsec:armsrace}
Even though hundreds of paper on bot detection gets published yearly, it needs to catch up to the malicious users. In other words, this results in an arms race between bot developers and bot detectors. We conjecture that the call to the arms race mainly started somewhere when the Oxford Internet Institute (OII) announced its strategy of detecting bots. According to OII, if an account posts over 50 tweets daily, it is a sign of automation, or ``heavy automation'' \cite{howard2016bots}. To keep the bot off the bot detection radar, a malicious user (bot developer) put in considerable effort to keep the tweet and retweet level below the threshold of 50 \cite{doi:10.1177/20539517211033566}. Since then, researchers have known that the attacker might game their solutions. 

Thus, looking at several more approaches, bot detection can be divided into standard features (hand-written or derived from OSN accounts) and robust features (evaluated for resistance to malicious manipulation). While the two approaches are not mutually exclusive, they differ in considering the attack and cost analysis of the extracted features. For example, Cresci et al. \cite{common7} used a standard feature of transforming profile timeline activity into a DNA-like string to detect coordinated spambots. At the same time, Xu et al. \cite{deepfacebook} employed deep features such as users in the photo, original creator, and users commenting, making feature exploitation costly for malicious users. The table in Appendix~\ref{appendix:feature_manipulation} shows the works that cover bot detection under different considerations. We conjecture that researchers widely use standard rather than robust features trained using adversarial training. 
\begin{tcolorbox}[colback=red!9!white, top=0pt, bottom=0pt, left=0pt, right=0pt]
\textbf{Research Gap 5. }
\textit{We notice that most works focus on simple rules while creating bot detection solutions. To improve the detection, robust features help against the sophisticated bot user. Machine learning and NLP processing are leveraged for sophisticated features~\cite{chang2022comparative,Ferrara_2023}.}
\end{tcolorbox}

\subsubsection{Bot Data Generation Methods}
Despite significant work on bot detection, the lack of large-scale benchmark datasets affected performance. To tackle the issue of the limited dataset, researchers resort to data generation techniques. These augmentation techniques, also called oversampling techniques, vary in the form of simulated data they produce, such as producing new data with high variance rather than simple copies of the original data. 
Researchers often use metrics such as the Kolmogorov-Smirnov test and Kullback-Leibler Divergence to evaluate the effectiveness of simulated bot data. Traditional oversampling techniques such as (SMOTE) \cite{10.1145/1007730.1007735} and (ADASYN)\cite{4633969} generate synthetic data with high metrics values, leading to the exploration of advanced oversampling methods such as (GA) \cite{cresci2019better}, (VAE) \cite{deng2020variational}, and (GANs) \cite{wu2019detecting, dialektakis2022caleb} to maintain acceptable thresholds. However, previous works assume one category's evolution and feed another's static data, resulting in high variability between both classes, potentially leading to incomplete or inaccurate evolution modeling.

\begin{tcolorbox}[colback=blue!9!white, top=0pt, bottom=0pt, left=0pt, right=0pt]
\textbf{Takeaway 5:}
    \textit{Research considering oversampling methods has used GAs, VAEs, and GANs to generate synthetic bot data due to low KS and KL values.} 
    \end{tcolorbox}

\subsubsection{Comment on Bot Research Datasets} 
Several researchers collaborated to create a Botometer repository aimed at facilitating bot research \cite{dimitriadis2021social}. While the repository contains human and bot profiles, limited research has been conducted on the bots' lifespan and replicability. It was discovered that the bot data in the repository is outdated and decreasing yearly, with only 40\% of the original data available for analysis \cite{dimitriadis2021social}. As of September 2022, there were only 66,348 active accounts out of the initial 174,306, highlighting significant data loss and replicability issues for future bot research studies. Additionally, bot detection tools rely solely on profile metadata which may be ineffective due to the volatility of such metadata~\cite{akhtar2022machine}.

Moreover, most of the existing research on bot detection is focused on using datasets by assuming that the label remains static throughout the data's lifespan, as is typical for other machine-learning tasks. However, bot dataset labels are volatile and subject to change due to various factors, such as the evolution of accounts and the changes to several metadata that come with it. A three-month study by Adrian et al. \cite{rauchfleisch2020false} demonstrated that bot scores fluctuate and remain unstable, raising an interesting question for research, which generally assumes a one-label-for-life approach to newly collected data. Due to Twitter policy issues, bot dataset owners can only share Twitter user IDs and class labels, requiring new researchers to recollect data from Twitter to reproduce results. However, this process may change the newly collected data compared to the original dataset for which labels were present.

\begin{tcolorbox}[colback=blue!9!white, top=0pt, bottom=0pt, left=0pt, right=0pt]
\textbf{Takeaway 6.}\textit{ Old bot labels do not belong fully to the new account metadata and newly collected timeline data.}
\end{tcolorbox}

\section{Third element of SMM: Malicious Campaigns}
\label{sec:campaign_detection}
OSN effectively helps businesses exchange content and reach new customers using technological penetration rather than traditional marketing \cite{campaign1}. Therefore, \textit{campaigns use coordinated efforts using social networks to maximize objectives and goals}\cite{almashor2021characterizing}. As such, Baum et al. \cite{campaign1} analyzed the impact of such campaigns on user actions and found that campaigns positively influence customer attitudes. Not only does it increase their intention to purchase, but it also recommends behavior. Since campaigns act as a business process, their performance can be measured using time series and sentiment data from the follower network \cite{campaignperformance}. However, there is a need for more complex campaign performance metrics to judge its performance. 

\textbf{Factors of Successful Campaigns.}
Due to the effectiveness of the campaigns, Sania et al. \cite{campaign2} studied the impact on people's attitudes and why marketing campaigns succeed. They used various variables such as \textit{vividness} (ability to evoke strong emotion), \textit{content of posts} (material used in advertisements), \textit{scheduling} (timing, frequency, and cycle of the campaign), and \textit{call to action} (immediate action required from the user) conducted on 300 respondents. Their research shows that the \textit{customer has a significant relationship with vividness, the content of posts, and scheduling, while the call to action has an insignificant effect} \cite{campaign2}.   

Despite ``content of posts'' being highlighted as a significant factor in~\cite{campaign2}, only a small proportion of messages outperform the rest. Therefore, Eismann et al.~\cite{campaign3} studied the essential drivers of Super Successful Posts (SSPs) based on likes, comments, and shares by taking an automotive industry-related 42 Facebook SSPs. They captured common patterns of SSP in five dimensions viz., ``co-branding, timing, cognitive task, wow effect and campaign''~\cite{campaign3}. Their work underlined the importance of \textit{`repeated exposure' of a message to increase the user's likelihood of engaging with the post}.

\begin{tcolorbox}[colback=blue!9!white, top=0pt, bottom=0pt, left=0pt, right=0pt]
\textbf{Takeaway 7.}\textit{ Repeated exposure of a message is an essential factor for the success of a campaign.}
\end{tcolorbox}

In recent times, malicious users are also aware of such campaigning techniques. In addition, malicious actors are using complex strategies to conduct malicious campaigns. For example, Daniel et al. \cite{ecuador} collected evidence from the 2017 Ecuador presidential election campaigns and found 32,672 bots, confirming the high use of political bots supporting candidates. 
Ferrara \cite{french} studied the disinformation campaigns of the 2017 French presidential election and found highly coordinated effort among users who shared patterns, with 18,324 social bots out of 99,378 users. 
Thus, it behoves us to study different mechanisms of malicious campaigns.

\subsection{Mechanism of Malicious Campaign}
\label{sub:mechanism}
{Much similar to the \textit{misinformation machine} defined by Ruths \cite{ruths2019misinformation} that views all the research in misinformation not against, but instead as a complementary and interconnected part of a more extensive system. Similarly, the\textit{ machinery of malicious campaigns} is a more significant part of numerous different research on strategies coordinated bots employ. Most works have focused on fragmented and specific-category \cite{vargas2020detection,boichak2021not,akhtar2022machine, Ferrara_2023}. For instance, during different events such as elections, hate speech, misinformation about healthcare, extremist ideologies, and financial market manipulation, bots have strategized and executed coordination to achieve their goals. In this sub-section, we will explore some of the interconnected processes of the global manipulation machine and realize the need for studying strategies and intent of malicious campaigns.} 
    
\textbf{Election and Paid Trolls.} 
{Zannettou et al., \cite{zannettou2019let} explored the evidence of state-sponsored actors that play a role in manipulating political events by using dedicated \textit{troll accounts}. In their work, Zannettou et al. analyzed Russian and Iranian trolls and found they were \textit{highly dependent upon URLs for sharing their propaganda}. In addition, authors analyzed the temporal activity of the troll accounts and found them to be \textit{active in the first hours of the day and first days of the week}.    

\textbf{Malicious Phishing URL campaigns.}
Almashor et al., \cite{almashor2021characterizing} studied malicious URLs that are typically used within phishing attacks for malware distribution. To deceive users, \textit{attackers impersonate brand URL} such as URLs resembling Apple or Paypal. Also, URLs used Transport Layer Security for downloading malware to users' devices, and very few of them were flagged by vendors. In addition, few malicious URL campaigns utilized fileless malware to evade detection methods. 

\textbf{Online Hate (Hate speech, toxicity, online abuse.)}
Qayyum et al. \cite{qayyum2022deep} identified toxic Twitter users by analyzing them from different perspectives. The methods vary from URL analysis, hashtag analysis, topics, readability, and homogeneity of domains to temporal activity. It is found that \textit{toxic users are more persistent in their activity and behave in an automated way}. Moreover, focused \textit{toxic profiles scores low on readability test} compared to random profiles.

\textbf{Disinformation campaigns.}
Vargas et al. \cite{vargas2020detection} studied the coordination of various strategic influence operations using network analysis. In their work, authors found \textit{network features such as co-retweet, co-hashtag, and co-URL useful}. Moreover, considering the scale of OSN, network analysis is expensive and complex. Yet, the network yields great results since \textit{depending on single coordination network lacks the generalized strategies} used by fake accounts.

\textbf{Extremist ideologies.}
With the onset of OSN, terrorist groups started conducting malicious activities such as sharing extremist propaganda or violent content. Almoqbel and Xu \cite{Almoqbel2019ComputationalMO} studied a strategy named \textit{`Twitterstorm,' which facilitated terrorist groups to trend their hashtags}. }

\begin{tcolorbox}[colback=blue!9!white, top=0pt, bottom=0pt, left=0pt, right=0pt]
\textbf{Takeaway 8. }\textit{While it is important to detect strategies behind bots and their campaign \cite{perspective}, we notice that the strategies change under different campaigns \cite{arif2018acting}.}
\end{tcolorbox}

\begin{center}
\begin{table*}[]
    \centering
\caption{Intent and Strategies identified using SoCIAL in malicious campaigns.}
\centering
  \scalebox{0.83}{
    \begin{tabular}{p{0.07\textwidth} p{0.20\textwidth} p{0.39\textwidth} p{0.18\textwidth} p{0.17\textwidth} p{0.07\textwidth}}
  \hline
  \hline
    \textbf{Ref.} & \textbf{Campaign Properties}	& \textbf{Explanation} & \textbf{Information Manoeuvre} (Strategies)& \textbf{Motive Identification} (Intent)& \textbf{Detection}\\ \hline
    \hline
    \cite{shao2018spread} & Bots active in initial false information spread & Bots post misinformation in the early phase to reach more people and induce them to share it. & Publicness, Accountability& Opportunity, Covertness& Easy \\\hline

    \cite{botcamp} & Bots mention organization/celebrity accounts & Bots mention a lot of organization and influencer users to get attention from them to reach more people. & Publicness& Integration& Easy \\\hline

    \cite{qayyum2022deep} & Bots don’t follow day and night cycle & Bots post at regular intervals without showing day/night cycle & Availability& Autonomy& Easy \\\hline

    \cite{howard2016bots,10.1145/3110025.3110090} & Bots lack original content and retweet more & Bots retweet more aggressively, do content theft, and humans generate more tweets & Use and Retention, Choice and Consent& Responsiveness, Covertness& Easy \\\hline

    \cite{10} & Bots maintain a balanced Follower/Followee ratio
 & Bots maintain a balanced ratio to evade detection tools. They adopt a strategy where one account tweets the original spam message, and the rest retweet it.
& Monitoring and Enforcement, Publiness& Covertness& Medium \\\hline

    \cite{shao2018spread} & Bots share more low-credibility content
& Bots share articles from low-credible sources rather than reputable outlets
& Integrity, Accuracy, Authenticity& Ethics, Collusion& Medium \\\hline

\cite{knight_foundation} & Bots share more conspiracy theories articles
& Bots share conspiracy theories, clickbait titles, and fake articles from fake websites. 
& Integrity, Accuracy, Accountability& Opportunity, Collaboration, Collusion& Medium \\\hline

\cite{alternative} & Bots promote ``alternative narratives'' of crisis events
& Bots share information by adding alternative narratives, for example, blaming other entities behind a cause
& Accountability, Authorization, Authenticity& Opportunity, Collusion& Complex \\\hline

\cite{KEIJZER2021100106} & Bots exert weak influence of opinion on neighbor & Using Axelrod’s seminal model (simulation), bots show that weak influence over neighbors is more effective in reaching opinions to farther neighbor than a strong influence on immediate neighbors. & Publicness, Integration& Covertness& Complex \\\hline

\cite{qayyum2022deep} & Bots take more part in online misbehavior & Focused profiles in Twitter Toxic Tweets take part in online toxicity and are automated in characteristics & Choice and Consent, Use and Retention& Ethics, Impersonation & Complex \\\hline

  \hline
  \hline
  \end{tabular}}
  \label{tab:campaign_complexity1}
  \end{table*}
\end{center}

\vspace{-2em}
\section{`SoCIAL' Framework for Assessing SMM} 
\label{appendix:social}
{
\textbf{Background}: In Section~\ref{sub:mechanism}, we highlighted the need for strategies (information maneuvers) and intent identification in malicious campaigns. Furthermore, our research indicates 
that strategies employed can vary across different campaigns. For example, 
Nimmo \cite{nimmo2015anatomy} identified four overarching tactics used in Russian influence operation: dismiss, distort, distract, and dismay. These tactics involve dismissing 
negative remarks, 
to manipulating data, creating distractions, and causing dismay. However, recent studies, such as \cite{carley2020social}, have shown that influence malicious campaigns are not limited to these tactics, emphasizing the necessity 
for a more detailed description of their intentions. 

On the other hand, campaigns in OSN not only deal with cybersecurity instead social cybersecurity as these directly impact humans on OSN. As Kathleen Carley \cite{carley2020social} stated, ``social cybersecurity is a field of an applied computational social science'' that characterizes and protects the social cyberinfrastructure. Thus, Carley proposed a BEND framework consisting of 16 maneuvers for analysis based on the micro level, such as the tweet-level information. 

\textbf{Working and Comparison:} Malicious campaign maneuvers involve more than just four strategies. Their identification using high-level campaign property instead of tweet-level gives more work flexibility. In contrast, we introduce the ``\textit{Social Cybersecurity: Campaigns, Information, and Actors Language }(SoCIAL)'' framework to assess which strategies and the intent are used in the malicious campaign from the social cybersecurity perspective. Therefore, drawing insights from existing research on social cybersecurity~\cite{obada2022sok,thomas2021sok}, we present strategies and intent (motive) into 29 social cybersecurity characteristics associated with four attack vectors. We have identified `eight' characteristics that malicious actors use to \textit{exploit} information to achieve malicious goals, `nine' behavioral characteristics that are abused and exercised by a malicious actor, `seven' characteristics of a direct attack on the user, and `two' characteristics of attacks on the OSN platform. 

\textbf{Impact:} The SoCIAL framework is a methodology used to assess which maneuvers are used in campaigns to inspire mitigation solutions against such maneuvers. For example, a particular campaign can be where bots mention influencers on OSN to get their post viral. When we assess such campaigns, we see the use of the ``publicness'' maneuver, such that bots want their post to be seen by the majority public on OSN. The intent behind such a campaign is ``integration'' to join other like-minded users on OSN. Based on this, OSN providers may re-evaluate who can mention who on OSN. In another example, bots maintain a balanced follower-to-following ratio. In this, bots show ``monitoring and enforcement'' maneuvers to escape defense mechanisms. Additionally, these bots display ``covertness'' as they try to achieve their goal without drawing the audience's attention. Detection of such a campaign in real-time is more challenging. 

In (Appendix Table~\ref{tab:Attack_vectors}), we show each characteristic with its meanings and the associated attack vector, such as whether the social cybersecurity characteristic can be used for exploitation (at the information level), abuse (at the behavior level) or attack on a user or platform. Consider this as an example, \textit{availablility} attack in traditional cybersecurity refers to a denial of service attack, whereas, in social cybersecurity, it refers to restricting a user from posting anything on OSN. Likewise, \textit{integrity} attack refers to every OSN account calling another ideologically dissimilar account as a bot. 
Similarly, we applied the SoCIAL framework on different campaign properties to identify intent and strategies as presented in Table~\ref{tab:campaign_complexity1}. 
We have two noteworthy observations based on our literature analysis (cf. Table~\ref{tab:campaign_complexity1}) on OSNs. Firstly, current works lack adequate solutions for addressing strategies (information maneuvers) and identifying intent in malicious campaigns~\cite{Ferrara_2023}. Secondly, we observe that strategies employed in the campaigns vary across different contexts. Our work calls for more interdisciplinary research to tackle the SMMs.} 

% \vspace{-0.3em}
\section{Open Issues and Future Trends}
\label{sec:openissues}

\subsection{No-large Scale Available Datasets on SMM}
In literature, available fake news datasets have varied characteristics. Most datasets are insufficient for building a robust machine-learning model due to the limited dataset size. Furthermore, existing datasets have covered various topics and domains of categories, although politics and entertainment are the most commonly explored. On the other hand, multimodal datasets are still limited in research. 

Similarly, bot detection research needs a large labeled bot dataset. A challenge to acquiring bot datasets from OSN such as Twitter is that bot profiles are suspended once detected. We see only 26\% of bots profiles remaining of the original Botometer dataset \cite{dimitriadis2021social}. 
Therefore, the availability of a large, diverse dataset is the need of the hour in research. 
On the other hand, GAN and VAE have the most potential for creating synthetic bot data to balance the different types of bot classes.

\subsection{Few Recovery Methods and Early Detection Techniques Available for SMM}
Given that a piece of information is misinformation, only a few methods exist to prevent dissemination. One method of preventing the widespread reach of false information is providing informative labels on posts to avoid users being misled in OSN. Such indicator, however, should not be restrictive but rather informative \cite{nutrition}. 
Furthermore, source ratings and flagging bot accounts with a caution label can be adopted.

On the other hand, even though the existing classifiers are known for achieving high accuracy in their proposed models, these models need to pay more attention to the urgency and need for early detection techniques. 
Moreover, much of the focus has been on evaluating the diffusion spread or the impact bots have on the eco-web of OSN. However, little light is extended on the strategies bots conduct and intent before carrying out the malicious campaigns \cite{9518390}. A potential venue for early detection of influence operations and malicious campaigns involves examining the relationship modeling and operational patterns of fake across various campaigns.

\subsection{High Requirement of Data for the Combined Study of SMM}
The detection of SMM involves identifying and monitoring the activities of bad actors, particularly those involved in propagating mis/disinformation. To achieve this, researchers often rely on annotated datasets of false information, which serve as ground truth in bot detection studies aimed at identifying malicious or bad actors disseminating false information. Additionally, researchers have discovered instances of bots-in-the-wild at OSNs.
However, to effectively detect malicious campaigns, it is important to have labeled data for both false information and bots.  
As such, identifying coordinated efforts among fake accounts,  
interpreting their intentions, and understanding their strategies becomes challenging 
without the availability of ground truth information, 
limiting in forming definitive conclusions. 
We also need more work on all three elements combined. 
So far, researchers have remained focused within their domains and need to incorporate interdisciplinary research. We also highlight that bots have been a critical element of SMM which has been neglected in SMM research.

\subsection{Is social media a critical infrastructure?}
Countries such as Australia define critical infrastructure as “those physical facilities or information technologies networks that, if destroyed, could significantly impact the nation's social and economic well-being” \cite{criticalInfrastructure}. Potential owners and operators manage critical infrastructure by analyzing risks and building resilience against malicious actors gaining illegal infrastructure control. However, such resiliency and risk assessments seem missing from the cyber realm attacks where the enemy could penetrate with minimal effort and achieve high success. Social media manipulations are silent weapons where a lousy actor can be just one friendship link away. Therefore, considering the cyber realm attacks such as social media manipulations, OSN should be part of the critical infrastructure, as a successful attack on this infrastructure could potentially lead to social and economic fallout for the nation.

\section{Conclusion}
\label{sec:conclusion}
Our research offers the first SoK to present the interconnection between each element of SMM by comprehensively discussing the OSN manipulation landscape. In doing so, we discuss existing defenses against each element of SMM and highlight various gaps and takeaways. We call for research to incorporate different interdisciplinary viewpoints due to the interplay between different SMM elements. We also briefly discussed ``SoCIAL'', a social cybersecurity framework that identifies the adversary's different intent and strategy in SMM by applying it to a few malicious campaigns. To this end, our systematization informs researchers and OSN admins that much further work is needed toward a safe OSN.

\bibliographystyle{ACM-Reference-Format}
% \bibliography{bib}

% % --- Appendix ---%
\appendix
% \section{Overflow form other sections}
% \label{sec:set-diff-dodis}
% Sometime you ware super excited about some details that does not quite fit with
% the rest of the paper goes here. For example, some details about how you
% instrumented the Android Linux kernel should go to appendix, and for really
% curious reader to read. Remember it's appendix, so the reader is not required to
% read, and you should not put critical information in appendix that is crucial
% for understanding the rest of the paper.

\section{Systematic Review Protocol Search keywords}
\label{appendix:systematic}
{As mentioned in Section~\ref{sec:protocol}, we retrieved literature by following a systematic review process. In this appendix, we explain the keywords used for the search process. First, we start by querying false information-related articles; we use the search terms with keywords such as ``detection" AND (``false information" OR ``fake news" OR ``misinformation" OR ``disinformation") AND (``machine learning" OR ``nlp") AND (``social media" OR ``online social networks" OR ``OSN"). Similarly, to find articles for bot detection, we use search query terms such as ``detection" AND (``bot" OR ``fake account" OR ``inauthentic account" OR ``automated") AND (``machine learning" OR ``nlp") AND (``social media" OR ``online social networks" OR ``OSN"). Likewise, to find articles for malicious campaigns, we use the keywords such as ``detection" AND (``malicious campaign" OR ``influence operation" OR ``computational propaganda" OR ``misinformation campaign" OR ``disinformation campaign") AND (``machine learning" OR ``nlp") AND (``social media" OR ``online social networks" OR ``OSN"). Finally, we combined all three forms and retrieved the final set of papers.} 

\section{Bots on Twitter}
\label{appendix:bot_twitter}
Different OSNs have different usability roles among users. For example, Twitter is considered one of the sources of information even for journalists. As discussed in Section~\ref{subsec:osnusers} concerning bots, on Twitter particularly, we identified bot percentages discovered from multiple studies in their datasets, ranging from 5\% to 25\% as shown in Table~\ref{tab:twitter_bot}. We note that Botometer is a widely used tool for bot detection with a threshold of 0.5.
\begin{table}[H]
\caption{Percentage of Bots on Twitter.}
\label{tab:twitter_bot}
\begin{center}
\scalebox{0.9}{
\begin{tabular}{c|c|c|c|c} 
  \hline
  \hline
      \textbf{Ref.} &	\textbf{Sample Size} & \textbf{Method} & \textbf{Threshold} & \textbf{\%age}\\
  \hline
  \hline
   \cite{us_sec_twitter} & >1000 (mDAU) & Twitter Inc. & N/A & 5\%-8.5\% \\ \hline
 \cite{shao2018spread} & 1000 & Botometer & >= 0.5 &6\%\\ 
  \hline
  \cite{usmidtermpaper} & 254,492 & Multiple Tools & >= 0.8 & 8\%\\
  \hline
  \cite{doi:10.1126/science.aap9559} & 2,725,269 & Botometer & >= 0.5 &13.2\%\\
  \hline
\cite{featureanalysis} & 14 million & Botometer & >= 0.43 &9\%-15\%\\
\hline
\cite{wright2018don} & 88 million & Manual Rules & >= 0.5 &21\%\\
\hline
\cite{fakenewsnet} & 10,000 & Botometer & >= 0.5 &22\%\\
\hline
\cite{keller2019social} & 1,476,700 & Botometer & >= 0.76 &5\%-25\%\\
\hline
\hline
\end{tabular}
}
\end{center}
\end{table}

\section{OSN Manipulation Landscape}
\label{appendix:osnlandscape}
In support of the discussion presented in Section~\ref{sub:osnmanipulation}, we identified OSN manipulation coverage in different studies and presented it in Table~\ref{tab:fi_osn}. The table highlights that the majority of the works do not consider other SMM elements, such as false information works do not focus on bots and vice versa. Moreover, works that focus on bot detection only extract sentiment features from the information posted by the user rather than verifying their falsity and providing a score. Lastly, we did not find a sufficient technical paper on malicious campaign detection to put in the table~\ref{tab:fi_osn}. The gap between different elements of SMM realizes that more interdisciplinary research against SMM is required. Works that focus on false information detection must also look into accounts propagating false information and their strategies of campaigns.

\begin{center}
\begin{table*}[t]
    \centering
\caption{Analysis of OSN landscape covered in the literature.}
 \centering
 \scalebox{0.82} {
  \begin{tabular}{l | c c c c c c c c c c | c c c c c}
  \hline
  \hline
    Ref. & \cite{doi:10.1126/science.aap9559} & \cite{doi:10.1080/10714421.2022.2129122} & \cite{fakeddit} & \cite{fakenewsnet} & \cite{stein2023realtime} & \cite{Liar} & \cite{EXIF} & \cite{UPFD} & \cite{EANN} &	\cite{areyoufakenews} & \cite{hayawi2022deeprobot} & \cite{subrahmanian2016darpa} & \cite{botcamp} & \cite{lstmdnn} & \cite{yang2020scalable}\\ \hline
    \hline
    \textbf{Information Reporting} & & & & & & & & & & & & & &\\ 
    \textit{Error} & \ding{55} & \ding{55} & \ding{55} & \ding{55} & \ding{55} & \ding{55} & \ding{55} & \ding{55} & \ding{55} & \ding{55} & \ding{55} & \ding{55} & \ding{55} & \ding{55} & \ding{55}\\ 
    \textit{Missing} & \ding{55} & \ding{55} & \ding{55} & \ding{55} & \ding{55} & \ding{55} & \ding{55} & \ding{55} & \ding{55} & \ding{55} & \ding{55} & \ding{55} & \ding{55} & \ding{55} & \ding{55}\\ 
    \textit{Biased} & \ding{55} & \ding{55} & \ding{55} & \ding{55} & \ding{55} & \ding{55} & \ding{55} & \ding{55} & \ding{55} & \ding{51} & \ding{55} & \ding{55} & \ding{55} & \ding{55} & \ding{55}\\ \hline

    \textbf{Event} & & & & & & & & & & & & & & &\\ 
    \textit{Random} & \ding{51} & \ding{51} & \ding{55} & \ding{51} & \ding{55} & \ding{55} & \ding{55} & \ding{55} & \ding{55} & \ding{55} & \ding{55} & \ding{55} & \ding{55} & \ding{55} & \ding{55}\\ 
    \textit{Predictive} & \ding{51} & \ding{55} & \ding{55} & \ding{55} & \ding{55} & \ding{55} & \ding{55} & \ding{55} & \ding{51} & \ding{55} & \ding{55} & \ding{55} & \ding{55} & \ding{55} & \ding{55}\\ 
    \textit{Time-critical} & \ding{51} & \ding{55} & \ding{55} & \ding{55} & \ding{55} & \ding{55} & \ding{55} & \ding{55} & \ding{55} & \ding{55} & \ding{55} & \ding{55} & \ding{55} & \ding{55} & \ding{55}\\ \hline

    \textbf{Intent} & & & & & & & & & & & & & & &\\ 
    \textit{Misinformation} & \ding{55} & \ding{51} & \ding{55} & \ding{51} & \ding{51} & \ding{55} & \ding{55} & \ding{55} & \ding{55} & \ding{55} & \ding{55} & \ding{55} & \ding{55} & \ding{55} & \ding{55}\\ 
    \textit{Disinformation} & \ding{55} & \ding{55} & \ding{51} & \ding{55} & \ding{55} & \ding{55} & \ding{55} & \ding{55} & \ding{55} & \ding{55} & \ding{55} & \ding{51} & \ding{55} & \ding{51} & \ding{55}\\ 
    \textit{Malinformation} & \ding{55} & \ding{55} & \ding{55} & \ding{55} & \ding{55} & \ding{55} & \ding{55} & \ding{55} & \ding{55} & \ding{55} & \ding{55} & \ding{55} & \ding{55} & \ding{55} & \ding{55}\\ \hline

    \textbf{Claim-Type} & & & & & & & & & & & & & & &\\ 
    \textit{Rumor} & \ding{51} & \ding{55} & \ding{55} & \ding{55} & \ding{55} & \ding{55} & \ding{55} & \ding{55} & \ding{55} & \ding{55} & \ding{51} & \ding{55} & \ding{55} & \ding{55} & \ding{55}\\ 
    \textit{Propaganda} & \ding{55} & \ding{55} & \ding{55} & \ding{55} & \ding{55} & \ding{55} & \ding{55} & \ding{55} & \ding{55} & \ding{55} & \ding{55} & \ding{55} & \ding{55} & \ding{55} & \ding{55}\\ 
    \textit{Fake News} & \ding{55} & \ding{55} & \ding{51} & \ding{51} & \ding{51} & \ding{51} & \ding{51} & \ding{51} & \ding{51} & \ding{55} & \ding{51} & \ding{55} & \ding{55} & \ding{51} & \ding{55}\\
    \textit{Spam} & \ding{55} & \ding{55} & \ding{55} & \ding{55} & \ding{55} & \ding{55} & \ding{55} & \ding{55} & \ding{55} & \ding{55} & \ding{51} & \ding{51} & \ding{55} & \ding{55} & \ding{55}\\ 
    \textit{Online Hate} & \ding{55} & \ding{51} & \ding{55} & \ding{55} & \ding{55} & \ding{55} & \ding{55} & \ding{55} & \ding{55} & \ding{55} & \ding{55} & \ding{55} & \ding{55} & \ding{55} & \ding{55}\\ \hline
    
    \textbf{Content context} & & & & & & & & & & & & & & &\\ 
    \textit{Satire} & \ding{55} & \ding{55} & \ding{55} & \ding{55} & \ding{55} & \ding{55} & \ding{55} & \ding{55} & \ding{55} & \ding{55} & \ding{55} & \ding{55} & \ding{55} & \ding{55} & \ding{55}\\ 
    \textit{False connection} & \ding{55} & \ding{55} & \ding{55} & \ding{55} & \ding{55} & \ding{55} & \ding{55} & \ding{55} & \ding{55} & \ding{55} & \ding{55} & \ding{55} & \ding{55} & \ding{55} & \ding{55}\\ 
    \textit{Misleading content} & \ding{55} & \ding{55} & \ding{55} & \ding{55} & \ding{55} & \ding{55} & \ding{51} & \ding{55} & \ding{55} & \ding{55} & \ding{55} & \ding{55} & \ding{55} & \ding{55} & \ding{55}\\
    \textit{False context} & \ding{55} & \ding{55} & \ding{55} & \ding{55} & \ding{55} & \ding{55} & \ding{55} & \ding{55} & \ding{55} & \ding{55} & \ding{55} & \ding{55} & \ding{55} & \ding{55} & \ding{55}\\ 
    \textit{Manipulated content} & \ding{55} & \ding{55} & \ding{55} & \ding{55} & \ding{55} & \ding{55} & \ding{55} & \ding{55} & \ding{55} & \ding{55} & \ding{55} & \ding{55} & \ding{55} & \ding{55} & \ding{55}\\ 
    \textit{Fabricated content} & \ding{55} & \ding{55} & \ding{55} & \ding{55} & \ding{55} & \ding{55} & \ding{55} & \ding{55} & \ding{55} & \ding{55} & \ding{55} & \ding{55} & \ding{55} & \ding{55} & \ding{55}\\  \hline

    \textbf{User Belief} & & & & & & & & & & & & & & &\\ 
    \textit{Confirmation Bias} & \ding{55} & \ding{55} & \ding{55} & \ding{55} & \ding{55} & \ding{55} & \ding{55} & \ding{51} & \ding{55} & \ding{55} & \ding{55} & \ding{55} & \ding{55} & \ding{55} & \ding{55}\\ 
    \textit{Naive Realism} & \ding{55} & \ding{55} & \ding{55} & \ding{55} & \ding{55} & \ding{55} & \ding{55} & \ding{51} & \ding{55} & \ding{55} & \ding{55} & \ding{55} & \ding{55} & \ding{55} & \ding{55}\\ 
    \textit{Homeostatis} & \ding{55} & \ding{55} & \ding{55} & \ding{55} & \ding{55} & \ding{55} & \ding{55} & \ding{55} & \ding{55} & \ding{55} & \ding{55} & \ding{55} & \ding{55} & \ding{55} & \ding{55}\\ 
    \textit{Homophily} & \ding{55} & \ding{55} & \ding{55} & \ding{55} & \ding{55} & \ding{55} & \ding{55} & \ding{55} & \ding{55} & \ding{55} & \ding{55} & \ding{55} & \ding{55} & \ding{55} & \ding{55}\\ \hline

    \textbf{Account Type} & & & & & & & & & & & & & & &\\ 
    \textit{Human account} & \ding{55} & \ding{55} & \ding{51} & \ding{55} & \ding{55} & \ding{55} & \ding{51} & \ding{51} & \ding{55} & \ding{55} & \ding{55} & \ding{55} & \ding{55} & \ding{55} & \ding{55}\\ 
    \textit{Bot account} & \ding{55} & \ding{55} & \ding{51} & \ding{55} & \ding{55} & \ding{55} & \ding{55} & \ding{55} & \ding{55} & \ding{55} & \ding{51} & \ding{51} & \ding{51} & \ding{51} & \ding{51}\\ 
    \textit{Cyborg account} & \ding{55} & \ding{55} & \ding{55} & \ding{55} & \ding{55} & \ding{55} & \ding{55} & \ding{55} & \ding{55} & \ding{55} & \ding{55} & \ding{55} & \ding{55} & \ding{55} & \ding{55}\\ \hline

    \textbf{Bot Type} & & & & & & & & & & & & & & &\\ 
    \textit{Social Bot} & \ding{55} & \ding{55} & \ding{51} & \ding{55} & \ding{55} & \ding{55} & \ding{55} & \ding{55} & \ding{55} & \ding{55} & \ding{51} & \ding{55} & \ding{55} & \ding{55} & \ding{51}\\ 
    \textit{Follower Bot} & \ding{55} & \ding{55} & \ding{55} & \ding{55} & \ding{55} & \ding{55} & \ding{55} & \ding{55} & \ding{55} & \ding{55} & \ding{55} & \ding{55} & \ding{55} & \ding{55} & \ding{55}\\ 
    \textit{Doppleganger Bot} & \ding{55} & \ding{55} & \ding{55} & \ding{55} & \ding{55} & \ding{55} & \ding{55} & \ding{55} & \ding{55} & \ding{55} & \ding{55} & \ding{55} & \ding{55} & \ding{55} & \ding{55}\\ 
    \textit{Political Bot} & \ding{55} & \ding{55} & \ding{55} & \ding{55} & \ding{55} & \ding{55} & \ding{55} & \ding{55} & \ding{55} & \ding{55} & \ding{55} & \ding{55} & \ding{55} & \ding{55} & \ding{55}\\ 
    \textit{Finance Bot} & \ding{55} & \ding{55} & \ding{55} & \ding{55} & \ding{55} & \ding{55} & \ding{55} & \ding{55} & \ding{55} & \ding{55} & \ding{55} & \ding{55} & \ding{55} & \ding{55} & \ding{55}\\ 
    \textit{Spam Bot} & \ding{55} & \ding{55} & \ding{55} & \ding{55} & \ding{55} & \ding{55} & \ding{55} & \ding{55} & \ding{55} & \ding{51} & \ding{51} & \ding{51} & \ding{51} & \ding{51} & \ding{51}\\ \hline

    \textbf{Campaign Promotion} & & & & & & & & & & & & & & &\\ 
    \textit{Organic} & \ding{55} & \ding{55} & \ding{55} & \ding{55} & \ding{55} & \ding{55} & \ding{55} & \ding{55} & \ding{55} & \ding{55} & \ding{55} & \ding{55} & \ding{55} & \ding{55} & \ding{55}\\ 
    \textit{Inorganic} & \ding{55} & \ding{55} & \ding{55} & \ding{55} & \ding{55} & \ding{55} & \ding{55} & \ding{55} & \ding{55} & \ding{55} & \ding{55} & \ding{55} & \ding{51} & \ding{55} & \ding{55}\\ \hline

    \textbf{Campaign Controller} & & & & & & & & & & & & & & &\\ 
    \textit{Single} & \ding{55} & \ding{55} & \ding{55} & \ding{55} & \ding{55} & \ding{55} & \ding{55} & \ding{55} & \ding{55} & \ding{55} & \ding{51} & \ding{55} & \ding{55} & \ding{55} & \ding{51}\\ 
    \textit{Group-based} & \ding{55} & \ding{55} & \ding{55} & \ding{55} & \ding{55} & \ding{55} & \ding{55} & \ding{55} & \ding{55} & \ding{55} & \ding{55} & \ding{51} & \ding{51} & \ding{51} & \ding{55}\\ \hline

  \hline
  \hline
  \end{tabular}}
  \label{tab:fi_osn}
  \end{table*}
\end{center}

\section{False and Real Information Diffusion}
\label{app:information_diffusion}
\begin{figure*}[h]
\centering
\subfloat[Real News Cumulative Count]{
\includegraphics[width=0.242\textwidth, keepaspectratio]{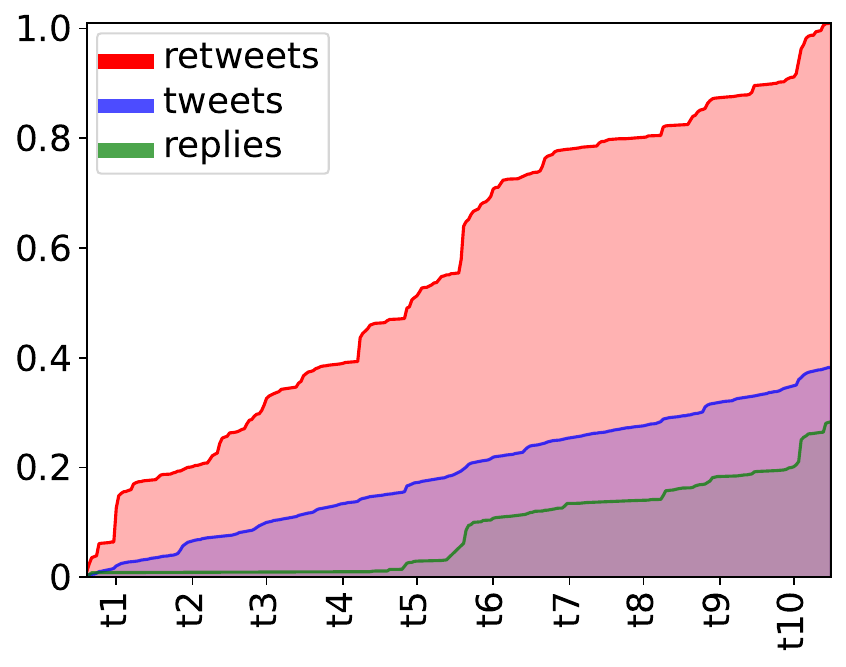} \label{app:real}}
\subfloat[Fake News Cumulative Count]{\includegraphics[width=0.247\textwidth, keepaspectratio]{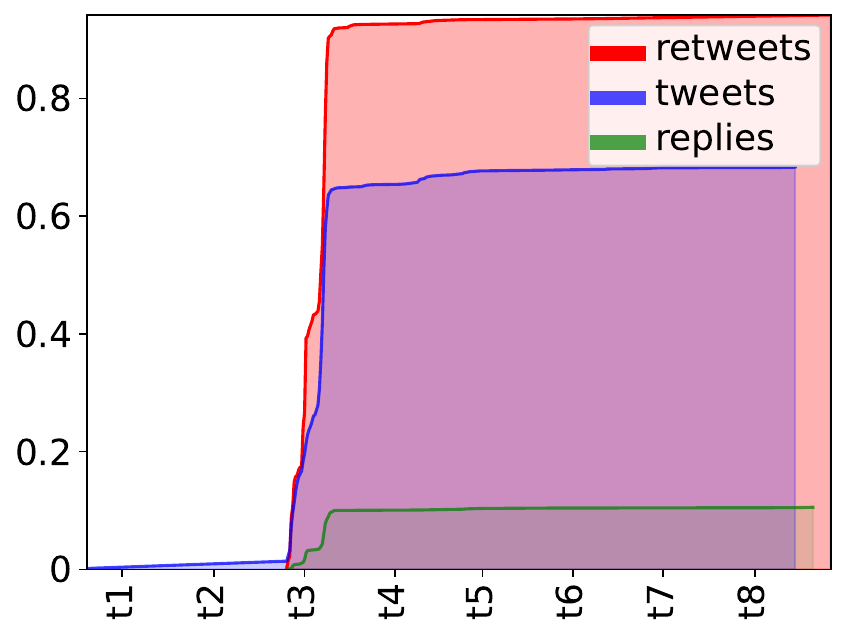} \label{app:fake}} 
\subfloat[Real News Spread Dist.]{
\includegraphics[width=0.237\textwidth, keepaspectratio]{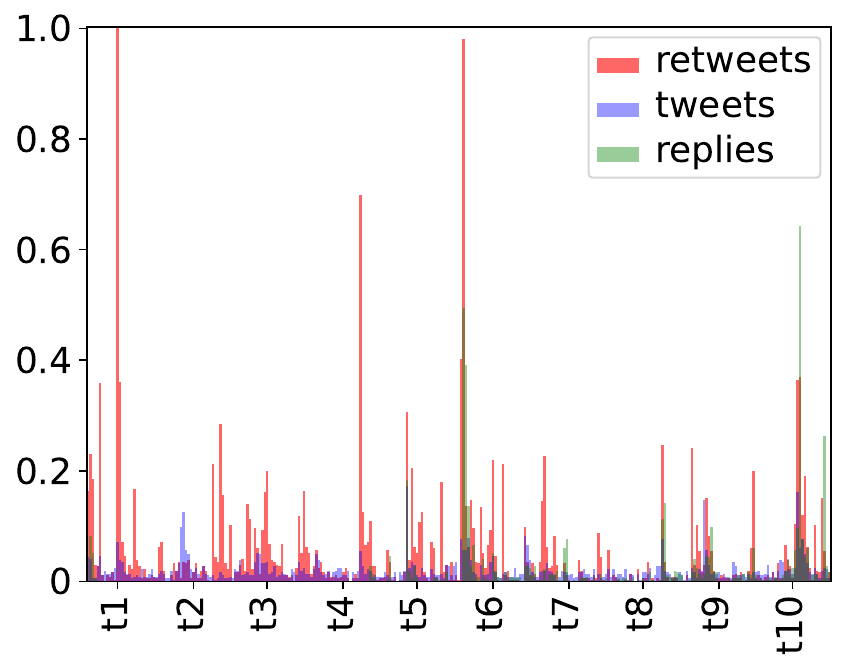} \label{app:real1}}
\subfloat[Fake News Spread Dist.]{\includegraphics[width=0.243\textwidth, keepaspectratio]{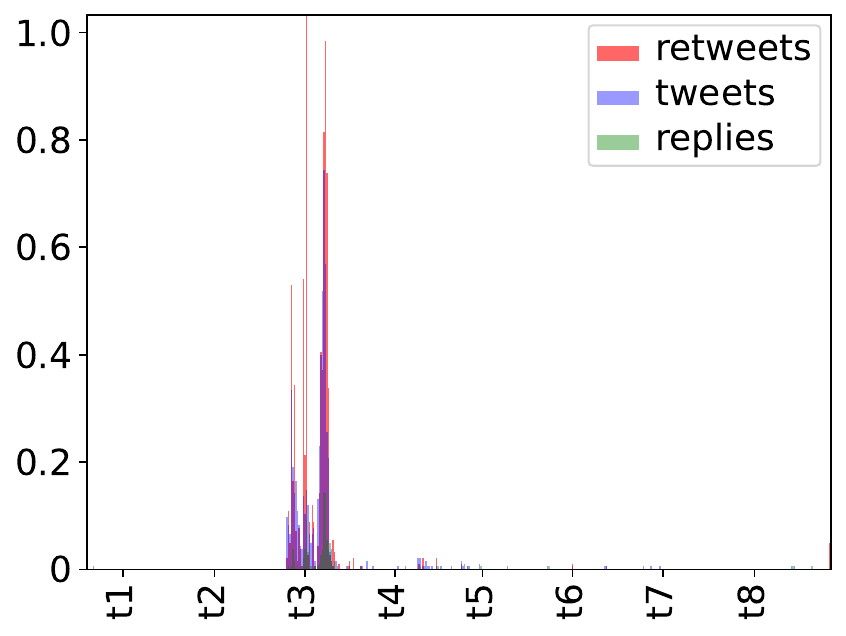}\label{app:fake2}}
\vspace{1em}
\caption{Real vs. Fake news' temporal normalized cumulative count and normalized distribution spread regarding tweets, retweets, and replies for Politics Category. Here, \textit{t1-t10} in real news refers to the time duration between 2018-03 to 2018-12, and \textit{t1-t8} in fake news refers to the time duration between 2016-07 to 2017-09.} 
\label{app:2}
\end{figure*}

Section~\ref{subsubsec:ddfi} delves into the characteristics of the diffusion dynamics associated with false information. While, in this section, we present our analysis of the cumulative distribution of false (or fake) and real information (or news) spread regarding tweets, retweets, and replies. %In 
Figure~\ref{app:2} depicts
the normalized cumulative count and distribution of these categories for both real and fake news. Notably, we observe that the fake information tends to generate a notable, sudden surge of retweets and tweets, surpassing the response garnered by the real information. 

\section{False information Characterization}
\label{appendix:characterization}
In Section~\ref{subsec:false_discussion} false information works are discussed with a literature review as shown in Table~\ref{tab:lit_works}. This appendix gives information on the characterization of literary work that is done using different collection methods, platforms, dataset sizes or languages, and modalities. To elaborate more in detail, the following characterization is presented:
\begin{itemize}
    
    \item \small \textbf{Data Collection Method}: Researchers have collected and annotated datasets manually (\fullcirc[0.8ex]) or gathered from existing data repositories (\emptycirc[0.8ex]) or a combination of both (\halfcirc[0.8ex]).
    
    \item \small \textbf{Platform}: Name of the platform for which the dataset is used.
    
    \item \small \textbf{Data Collection Tier}: It involves the level of data collection. For instance, \textit{Tier 0} (\emptycirc[0.7ex] \emptycirc[0.7ex] \emptycirc[0.7ex]): a) Only tweet text content or b) Tweet text and image, or c) News text content only, d) News text and image; \textit{Tier 1} (\fullcirc[0.8ex] \emptycirc[0.7ex] \emptycirc[0.7ex]): a) Tweet + account information or b) News information + meta-data (speaker, context, party affiliations), c) Images data only + meta-data, or d) Image + text in image + meta-data; \textit{Tier 2} (\fullcirc[0.8ex] \fullcirc[0.8ex] \emptycirc[0.7ex]): Account information + timeline data; \textit{Tier 3} (\fullcirc[0.8ex] \fullcirc[0.8ex] \fullcirc[0.8ex]): Account information + Timeline data + friends timeline
    
    \item \small \textbf{Size}: Count of the dataset size.
    
    \item \small \textbf{Classes}: Number of class instance and labels. Binary involves two classes whereas multi-classification problems consist of three or more classes.
    
    \item \small \textbf{Availability}: Whether code is available (\ding{73}) or data is available (\fullcirc[0.8ex]) or both (\ding{74}). Generally, due to ethics and privacy concerns, multiple researchers do not outsource datasets).
    
    \item \small \textbf{Feature and Model Selection (FS + Model)}: How features were extracted and model was build.
    
    \item \small \textbf{Popularity}: Code popularity on Github with respect to repository stars. This demonstrates that the work is useful and influential.
    
    \item \small \textbf{Duration}: This refers to temporal information when the dataset was collected. 
    
    \item \small \textbf{Domain}: Category of false information such as Politics, entertainment, etc. 
     
    \item \small \textbf{Language}: Dominant language used in the dataset on which detection is conducted.
    \item \small \textbf{Modality}: Type of content i.e., Unimodal (\ding{46}, text-based or image-based) or multimodal (\faCamera, text and image-based).
   
\end{itemize}

\section{Bots Feature Description}
\label{appendix:features}
Both inferential and descriptive approaches, as shown in Section~\ref{sec:bot_detection}, use various features in their model. We find five dimensions of feature groups for every account: user profile-based, content-based, temporal-based, devices-based, and network-based features. 
However, we have combined the device features with the user and temporal features with the network features, as one can be derived from the other.
 
\subsection{User Profile Features}
User features include meta-data features such as the number of tweets, number of followers/following, account creation date, verified status, and a few others, like short bio descriptions. These features are simple and range from numerical and binary to textual. As such, Beskow et al. \cite{beskow2018bot} used 6 user features, Hayawi et al. \cite{hayawi2022deeprobot} used 12 features, Yang et al. used 8 direct meta-data user features and 12 derived features, Kudugunta et al. \cite{lstmdnn} used 10 account-level features (such as status count, list count) and 6 tweet-level features (such as the number of hashtags, URLs) to distinguish bots. On another end, Botometer tool \cite{botornot} uses up to 1200 features to determine bot likelihood. 

Furthermore, these features can be trained with a supervised classifier like a Random forest model. We mentioned Random Forest, which has been widely adopted and accepted in several user feature-based bot detection researches \cite{latah2020detection}. Since ``\textit{Random forest model can learn nonlinear decision boundaries, can handle a large amount of dataset and unnormalized features for training}'' \cite{vargas2020detection}. However, due to the simplicity of the features, such models face the issue of feature tampering by adversaries. For example, maintaining a balance between followers and the following level takes less effort by a malicious user.

\subsection{Network Features} 
Twitter conversations are a combination of tweets and retweets. Ross and Andrew \cite{usmidtermpaper} analyzed that retweet networks can provide engagement activity between bots and humans and intra-group interactions. Less than 8\% accounts were bot among the total unique accounts. However, more than 20\% were bots in the top 100 and top-25 out-degree centrality rankings. This implies the persistent activity of bots to engage with humans irrespective of being small in population size. 

Moving ahead, Varol et al. \cite{featureanalysis} uses various network features such as network density, clustering coefficient, retweet network, and in-strength and out-strength (weighted degree) distribution. Such features help anomaly detection and provide signals for looking further into dense networks of followers and friends of suspicious accounts. Other recent works focus on graph-based techniques using network features, such as node centrality \cite{dehghan2022detecting} or Graph neural networks \cite{feng2022twibot}. However, graph-based techniques and network features are computationally expensive in terms of collection time. Collecting large-scale datasets from Twitter is a technical hindrance due to Twitter rate limit issues \cite{wright2018don, rate_limit}. According to Beskow and Carley \cite{beskow2018bot}, it takes about 20 hours to collect network information of 250 accounts. Thus, most of the work remains till Tier 2 of data collection, which consists of user and timeline information.

\begin{tcolorbox}[colback=blue!9!white, top=0pt, bottom=0pt, left=0pt, right=0pt]
\textbf{Takeaway A1.}\textit{ Network features are paramount for detection, though researchers avoid it due to data collection time and resource constraints.}
\end{tcolorbox}

\subsection{Content and Language Features}
Bot accounts post malicious URLs, misinformation content, and malware on OSN \cite{common1}. Thus, content information brings many contexts to detect bots. Even though different bots post on different topics, tweet-based bot detection uses similar features such as tweet text, tweet length, number of hashtags, URLs, mentions used, is\_possibly\_sensitive, similarity between tweets, the sentiment of the tweet, originality along with the user-metadata \cite{precisionandrecall}. On another end, Heidari and Jones \ proposed another solution \cite{9298158} to use sentiment features from content to detect bots, as bots purposely may be skewed to a position on a topic. However, we noticed that most works focus on user metadata and neglect the content information for social bot detection. 

\begin{tcolorbox}[colback=red!9!white, top=0pt, bottom=0pt, left=0pt, right=0pt]
\textbf{Research Gap A1.}\textit{ We note that limited work leverage features from content posted by bots. This may be an important feature to differentiate between benign and malicious bots.}
\end{tcolorbox}

\section{Case Study: Analysis of coordinated crypto-scams (example of \textit{Trigger} bots)}
\label{appendix:case_study}
As mentioned in Section~\ref{subsec:descriptive}, researchers often use the descriptive approach where clusters help to analyze the activity of accounts. Moreover, the descriptive approach also helps to verify claim about certain behavior. One such activity our team addresses in this appendix section is related to Trigger bots, a new family of spambots on Twitter. To aid with a piece of background knowledge, a frequent assumption is that humans are inefficient in detecting social spambots. Despite this, on Twitter, our team observed a few genuine users claiming they found bots. We further looked in a few other conversations and found genuine accounts have expressed their concern about bots that flock to the comment section when a certain \textit{magic words} are used on Twitter. 
We call such bots trigger bots. For example, one of the users wrote the word ``\textit{Metamask}’' (a famous cryptocurrency wallet) and 
expressed that many bots replied after seeing the word ``Metamask'' due to their automated crawling behavior. Interestingly, few bots replied by sharing a malicious phishing link. 

Following the analysis of user observations, a viability test was conducted. First, we tweeted by using the word ``Metamask'' in a sentence and expressed that we needed help with our wallet to provide some auxiliary context in the tweet. Within the first 10 hours, nine replies were received, and upon checking the Botometer score of each profile, it was found that most of them had a score of more than 4 on a scale of 0-5, indicating a strong possibility of a bot-like account. In the second test, a tweet containing only the word ``Metamask'' was posted without any additional context. The first reply was received in under nine seconds, and two more were received in less than seven minutes. One of the replies suggested we contact a phishing email address for account recovery, as they had experienced a similar issue. Given the lack of context in the tweet, it was hypothesized that bots were responding using a crawling method triggered by the magic words. (See Table~\ref{tab:twitter_bot} for more potential trigger words.)

\begin{figure*}[!ht]
\centering
\subfloat[]{
\includegraphics[width=0.22\textwidth, keepaspectratio]{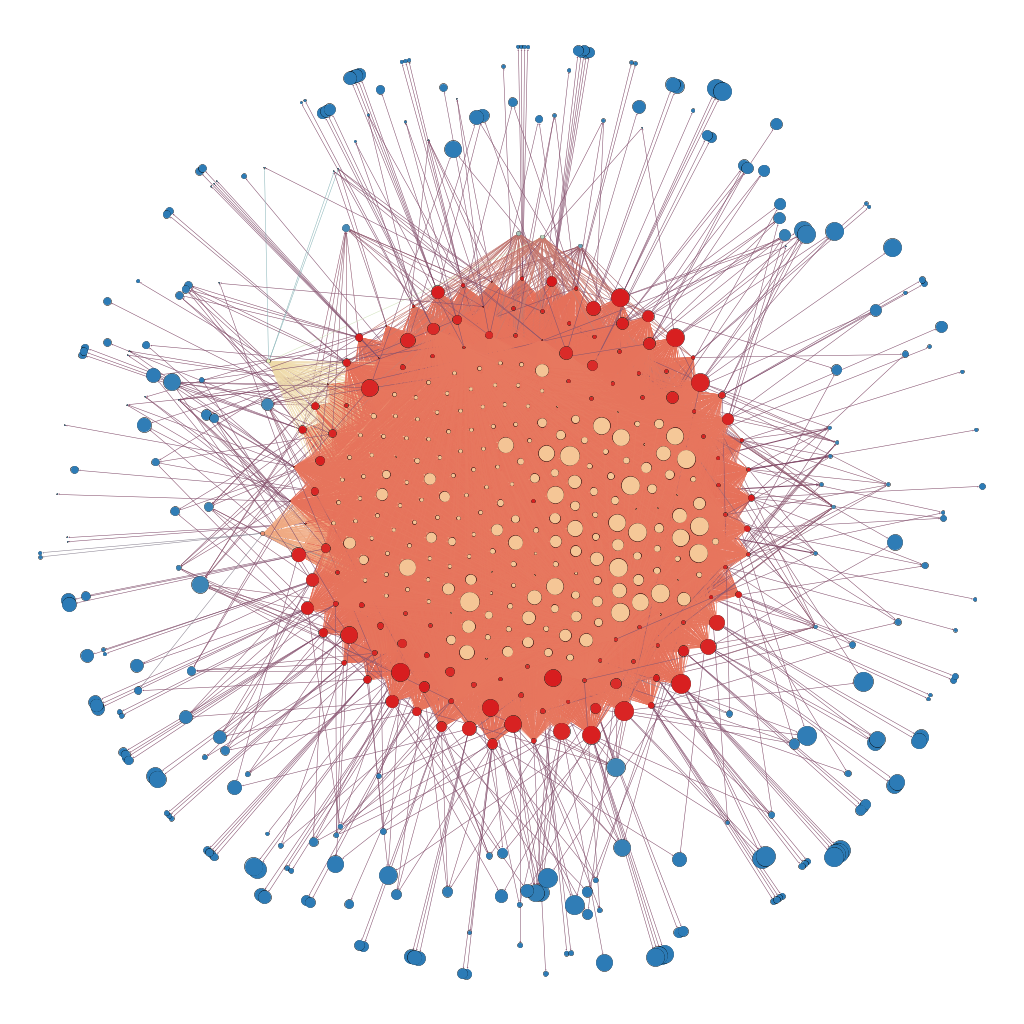} \label{fig:casea}}
\subfloat[]{\includegraphics[width=0.23\textwidth, keepaspectratio]{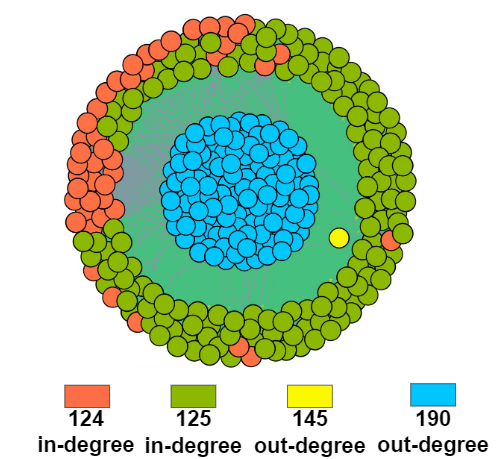} \label{fig:caseb}}
\subfloat[]{
\includegraphics[width=0.22\textwidth, keepaspectratio]{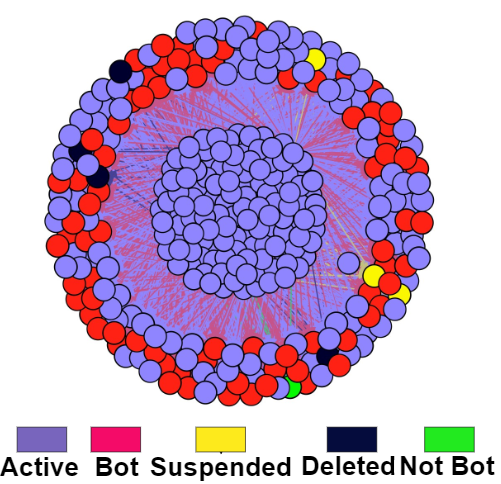} \label{fig:casec}} 
\subfloat[]{\includegraphics[width=0.22\textwidth, keepaspectratio]{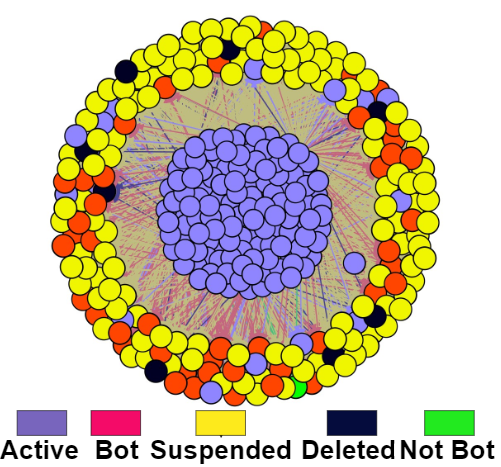} \label{fig:cased}}
\vspace{1em}

\caption{\small Visualization of crypto-scams (trigger) bots as depicted by (\protect{\ref{fig:casea}}) dense network of bots in orange color interacting with genuine users shown in blue color. On examining the orange network of trigger bots, we identified two concentric circles of accounts exhibiting two different roles---inner circle bots only tasked to follow outer circle bots and outer circle bots responsible for posting scam messages. These bots have a similar (in and out) degree metric as depicted in (\protect{\ref{fig:caseb}}); in this network of bots, (\protect{\ref{fig:casec}}) presents different accounts shown as humans, bots, suspended using Botometer tool as of \textit{06/07/2022}; and (\protect{\ref{fig:cased}}) different accounts shown as humans, bots, suspended using Botometer tool as of \textit{30/08/2022}. We used two different dates to show that many of the accounts from the outer circle are now suspended, but none of the accounts in the inner circle are suspended due to no timeline post activity.}
\label{fig:case1}
\end{figure*}

\begin{figure*}[!ht]
\centering
\subfloat[]{
\includegraphics[width=0.44\textwidth, keepaspectratio]{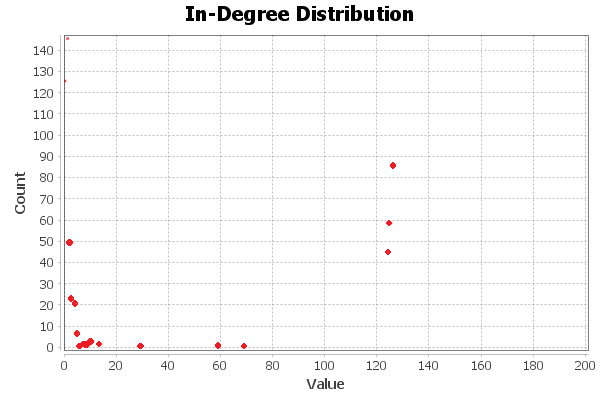} \label{fig:indegree}}
\subfloat[]{\includegraphics[width=0.44\textwidth, keepaspectratio]{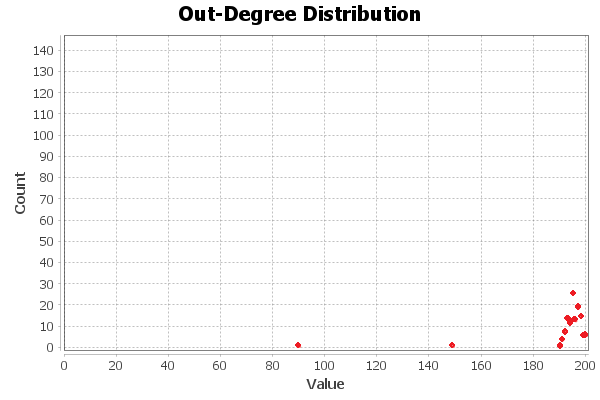} \label{fig:outdegree}}
\vspace{1em}
\caption{\small (\protect{\ref{fig:indegree}}) shows the in-degree distribution of outer circle trigger bots with many of the nodes with similar in-degree (close to 124)---present in the center of the figure; and (\protect{\ref{fig:outdegree}}) shows the out-degree distribution of inner circle of trigger bots with many bots having similar and high value of out-degree (close to 190)---present at the end of the figure.}
\label{fig:in_out_degree}
\end{figure*}

Next, we aimed to explore the trigger bot network since these accounts had an image in their profiles and few followers, mimicking real profiles. For this, we followed two steps. First, we selected a trigger bot seed account. Second, we build a network of the seed bot's followers and their respective friends (followings). For visualization, we used the Gephi tool\cite{gephi}. As shown in Figure \ref{fig:casea}, the dense pathogenic network in orange seems to follow the genuine users (users whom Twitter verified) seen in blue. Then, we filtered out the genuine users and used the Gephi visualization tool (with the Yifan Hu model) to dive deep into this dense network. We find it interesting to see that it resulted in two concentric circles with varied characteristics. First, the accounts in the inner circle are explicitly tasked to follow the accounts in the outer circle, making the outer accounts look popular and credible. We believe this is why every account in the inner circle had zero incoming links and 190 outgoing links. For the accounts in the outer circle, a few had 124 incoming links, and others had 125 incoming links, as shown in Figure \ref{fig:caseb}. Moreover, Figure~\ref{fig:in_out_degree} shows the in and out-degree distribution of accounts in both circles.

To give context, we found that only the outer profiles were used for malicious activities, such as posting phishing URLs and carrying crypto scams. \textit{These accounts were more persistent, visible, and tweeted with high responsiveness.} Then, we use the Botometer service\cite{botornot} to identify all the bot accounts in our two circles. We note that Botometer could only find bots from the outer circle. The accounts in the inner circle never posted anything, producing no timeline activity for detection. The bot accounts are represented in red in Figure \ref{fig:casec}. As of 6th July 2022, we see only three accounts suspended by Twitter, as shown in yellow in Figure \ref{fig:casec}. Whereas a month later, Twitter discovered and suspended most bot accounts from the outer circle, as shown in Figure \ref{fig:cased}. 
This corroborates our findings about the crypto-scams trigger bots. However, the inner circle accounts were still active since they did not post anything. These accounts could be used for any other purposes as well. This highlights the need for a relationship between bots and the malicious activities they conduct to detect accurately.

\begin{table}[H]
\caption{Trigger bots potential triggering keywords.}
\label{app:A_trigger_bots}
\centering
\scalebox{0.98}{
\begin{tabular}{p{0.6\linewidth}|p{0.28\linewidth}} 
  \hline
  \hline
\textbf{Sentence} &	\textbf{Potential Trigger Word}\\
  \hline
  \hline
  metamask & metamask\\\hline
  got hacked on Instagram & hacked \\ 
  \hline
  need someone to help me write this essay & essay \\
  \hline
  trust wallet & trust wallet \\
  \hline
my account got disabled & account disabled \\
  \hline
    need logo school & logo \\
  \hline
    help need banner  & banner \\
  \hline
    need design & design \\
  \hline
    need gfx & gfx \\
  \hline
  bot cashapp venmo & cashapp, venmo \\\hline
i need a sugar daddy, sugar daddy cheated & sugar daddy \\\hline
need this on a tshirt & tshirt \\\hline
password managers & password managers \\\hline
mushrooms & shrooms \\\hline
account suspended & suspended \\\hline

\hline
\hline
\end{tabular}
}
\end{table}

\begin{tcolorbox}[colback=blue!9!white, top=0pt, bottom=0pt, left=0pt, right=0pt]
\textbf{Takeaway A2.}\textit{We explored a new type of bots, named Trigger bots whose behavior gets triggered on certain magic words.}
\end{tcolorbox}
\setlength{\floatsep}{5pt plus 2pt minus 2pt}

\section{Bot detection and feature manipulation}
\label{appendix:feature_manipulation}
This appendix section describes the different considerations for features used in bot detection. As already discussed in sub-section~\ref{subsec:armsrace}, bot developers leverage the advantage of adversarial attacks on bot detectors. As shown in Table~\ref{tab:bot_detection}, most of the works use standard features instead of robust features. Hence, the bot detector remains vulnerable to adversarial attacks.

\begin{table}[H]
\caption{Bot Detection Works with Respect to Features.}
\label{tab:bot_detection}
\begin{center}
\scalebox{0.98}{
\begin{tabular}{p{0.23\textwidth}|p{0.20\textwidth}} 
  \hline
  \hline
      \textbf{Work Type} & \textbf{Ref.} \\
  \hline
  \hline
  Bot Detection with Standard Feature & \cite{walt_eloff,dna_modeling_cresci,common7,featureanalysis,sayyadiharikandeh2020detection,10.1093/cybsec/tyac015} \\\hline
  Bot Detection with Robust Feature & \cite{deepfacebook,9093747} \\\hline
  Bot Detection with Adversarial training, on Standard Feature & \cite{adverarial_cost_shao,cresci2019better,social_bots_fire,cresci2021coming} \\
\hline
\hline
\end{tabular}
}
\end{center}
\end{table}

\section{`SoCIAL' Framework for Assessing SMM} 
\label{appendix:social}
{Note that OSN campaigns not only encompass cybersecurity concerns but also have significant implications for social cybersecurity, directly impacting individuals on these platforms. 
{To address these gaps,} we introduce the ``\textit{Social Cybersecurity: Campaigns, Information, and Actors Language }(SoCIAL)'' framework. The primary objective of this framework is to assess which strategies and intent are used in malicious campaigns from the social cybersecurity perspective. 
Drawing insights from existing research on social cybersecurity~\cite{obada2022sok,thomas2021sok}, we systemize strategies and intent (motive) into 29 social cybersecurity characteristics associated with four attack vectors. 
These characteristics encompass eight aspects related to the \textit{exploitation} of information for malicious purposes, nine behavioral characteristics that are manipulated and utilized by malicious actors, seven characteristics of direct user-targeted attacks, and two characteristics of attacks on the OSN platform. 
Appendix Table~\ref{tab:Attack_vectors} provides additional information about the SoCIAL and its practical application.}

 \begin{center}
\begin{table*}[]
    \centering
\caption{\textbf{So}cial Cybersecurity: \textbf{C}ampaigns, \textbf{I}nformation and \textbf{A}ctors \textbf{L}anguage \textbf{(SoCIAL)} Framework.}
\centering
\scalebox{0.95}{
    \begin{tabular}{p{0.10\textwidth} p{0.30\textwidth} p{0.30\textwidth} p{0.14\textwidth}}
  \hline
  \hline
    \textbf{Principles/ Characteristics} &\textbf{Cybersecurity} (at machine and information perspective) & \textbf{Social Cybersecurity} (at human perspective on OSN) & \textbf{Attack Vector}\\ \hline
    \hline
    \textit{Integrity} & Actions that information remains unchanged  & Actions that reputation remains unchanged & Exploitation\\ 

    \textit{Authentication} & Actions that system able to verify and that data originated from its purported source  & Actions that users able to verify themselves and that data originated from its purported source & Exploitation\\ 

    \textit{Authorization} & Actions that determine user’s access right for a system resource & Actions that determine user’s right over information content & Exploitation\\ 

    \textit{Accuracy} & Action that deals with correctness of data & Action that deals with correctness of data & Exploitation\\ 

    \textit{Accountability} & Actions that ensure who is responsible for safeguard & Actions that ensure who is responsible  & Exploitation\\ 

    \textit{Use and Retention} & Actions that ensure data use and retention is as outlined or in contract & Actions that ensure data use and retention is as outlined or in contract & Exploitation\\ 

    \textit{Choice and Consent} &Actions that ensure system or user has permission to the collection, use, disclosure of data & Actions that ensure system or user has permission to the collection, use, disclosure of data & Exploitation\\ 

    \textit{Confidentiality} & Actions that protect data from unauthorized access and misuse & Actions that protect data from unauthorized access and misuse & Exploitation\\ \hline 

    \textit{Responsiveness} & - & Intent to ensure user is responsive  & Abuse\\ 

    \textit{Anonymity} & Intent to hide true identity & Intent to hide true identity & Abuse\\ 

    \textit{Evolution} & Intent to update the state of system & Intent to evolve user account behavior & Abuse\\

    \textit{Covertness} & - & Intent to achieve goal without exposing or drawing audience attention & Abuse\\ 

    \textit{Malleability} & - & Intent to associate in other activities & Abuse\\ 

    \textit{Integration} & - & Intent to join other like-minded user & Abuse\\ 

    \textit{Influence} & - & Intent to influence user & Abuse\\ 

    \textit{Publicness} & - & Intent to be seen by public & Abuse\\ 

    \textit{Engagement} & - & Intent to engage with public & Abuse\\ 
    
     \textit{Monitoring and Enforcement} & Intent to escape defence mechanism & Intent to escape platform policies and defence mechanism & Abuse\\ \hline

    \textit{Impersonation} & Attempt to create Sybil identities & Attempt to create fake profiles & Attack (User)\\

    \textit{Awareness} & - & Attempt to increase user awareness & Attack (User)\\ 

    \textit{Collusion} & - & Attempt to confuse user to make uninformed decision & Attack (User)\\ 

    \textit{Isolation} & Attempt to make systems run isolated for better security and minimal shared resources & Attempt to make the user feel isolated to make targeted attack & Attack (User)\\ 

    \textit{Availability} & Attempt to make systems unavailable for users to use & Attempt to make users unavailable to use systems & Attack (User)\\ 

    \textit{Reputation} & - & Attempt to alter user reputation & Attack (User)\\ 

    \textit{Minimization} & - & Attempt to counter manipulation minimization techniques & Attack (User)\\ \hline

    \textit{Proportionality} & - & Attempt to make user feel safeguard schemes are worthless & Attack (Platform)\\ 

    \textit{Ethics} & Attempt to violate ethics & Attempt to violate ethics & Attack (Platform)\\ \hline 

    \textit{Autonomy} & Mechanism to control the system & Mechanism to control the system  & Channels\\ 

    \textit{Opportunity} & - & Mechanism to participate opportunistically & Channels\\ \hline

  \hline
  \hline
  \end{tabular}}
  \label{tab:Attack_vectors}
  \end{table*}
\end{center}

%%% Local Variables:
%%% mode: latex
%%% TeX-master: "main"
%%% End:

\end{document}